\documentclass[twocolumn]{aastex631}

\usepackage{multirow}
\usepackage{subfig}

\shorttitle{Kinematically Identifying Stars from Ultra-Faint Dwarf Galaxies}
\shortauthors{Brauer et al.}

\graphicspath{{./}{figures/}}

\begin{document}

\title{Possibilities and Limitations of Kinematically Identifying Stars from Accreted Ultra-Faint Dwarf Galaxies}

\correspondingauthor{Kaley Brauer}
\email{kbrauer@mit.edu}

\author[0000-0002-8810-858X]{Kaley Brauer}
\affiliation{Department of Physics and Kavli Institute for Astrophysics and Space Research, Massachusetts Institute of Technology, Cambridge, MA 02139, USA}

\author{Hillary Diane Andales}
\affiliation{Department of Physics and Kavli Institute for Astrophysics and Space Research, Massachusetts Institute of Technology, Cambridge, MA 02139, USA}

\author[0000-0002-4863-8842]{Alexander~P.~Ji}
\affiliation{Department of Astronomy and Astrophysics, University of Chicago, Chicago IL 60637, USA}

\author[0000-0002-2139-7145]{Anna Frebel}
\affiliation{Department of Physics and Kavli Institute for Astrophysics and Space Research, Massachusetts Institute of Technology, Cambridge, MA 02139, USA}

\author[0000-0001-9178-3992]{Mohammad K.\ Mardini}
\affiliation{Kavli IPMU (WPI), UTIAS, The University of Tokyo, Kashiwa, Chiba 277-8583, Japan}
\affiliation{Institute for AI and Beyond, The University of Tokyo 7-3-1 Hongo, Bunkyo-ku, Tokyo 113-8655, Japan}

\author[0000-0002-1947-333X]{Facundo A. G\'omez}
\affiliation{Instituto de Investigaci\'on Multidisciplinar en Ciencia y Tecnolog\'ia, Universidad de La Serena, Ra\'ul Bitr\'an 1305, La Serena, Chile}
\affiliation{Departamento de F\'isica y Astronom\'ia, Universidad de La Serena, Av. Juan Cisternas 1200 N, La Serena, Chile}

\author[0000-0002-2786-0348]{Brian W. O'Shea}
\affiliation{Department of Computational Mathematics, Science and Engineering, Michigan State University, MI, 48824, USA}
\affiliation{Department of Physics and Astronomy, Michigan State University, MI, 48824, USA}
\affiliation{Facility for Rare Isotope Beams, Michigan State University, MI, 48824, USA}

\begin{abstract}

The Milky Way has accreted many ultra-faint dwarf galaxies (UFDs), and stars from these galaxies can be found throughout our Galaxy today. Studying these stars provides insight into galaxy formation and early chemical enrichment, but identifying them is difficult. Clustering stellar dynamics in 4D phase space ($E$, $L_z$, $J_r$, $J_z$) is one method of identifying accreted structure which is currently being utilized in the search for accreted UFDs. We produce 32 simulated stellar halos using particle tagging with the \textit{Caterpillar} simulation suite and thoroughly test the abilities of different clustering algorithms to recover tidally disrupted UFD remnants. We perform over 10,000 clustering runs, testing seven clustering algorithms, roughly twenty hyperparameter choices per algorithm, and six different types of data sets each with up to 32 simulated samples. Of the seven algorithms, HDBSCAN most consistently balances UFD recovery rates and cluster realness rates. We find that even in highly idealized cases, the vast majority of clusters found by clustering algorithms do not correspond to real accreted UFD remnants and we can generally only recover $6\%$ of UFDs remnants at best. These results focus exclusively on groups of stars from UFDs, which have weak dynamic signatures compared to the background of other stars. The recoverable UFD remnants are those that accreted recently, $z_{\text{accretion}}\lesssim 0.5$. Based on these results, we make recommendations to help guide the search for dynamically-linked clusters of UFD stars in observational data. We find that real clusters generally have higher median energy and $J_r$, providing a way to help identify real vs. fake clusters. We also recommend incorporating chemical tagging as a way to improve clustering results. 

\end{abstract}



\section{Introduction} \label{sec:intro}

Throughout its formation history over billions of years, the Milky Way grew through mergers with many dwarf galaxies. The smallest and oldest of these accreted systems are the ultra-faint dwarf galaxies (UFDs), which were among the first galaxies in the Universe \citep{frebel10,Simon19}. These systems provide insight into the earliest stages of galaxy formation and are important components of the assembly history of the Milky Way. 

Due to low star formation efficiency and quenching from reionization, UFDs preserve information about early chemical enrichment and can display clean signatures of important nucleosynthetic processes such as the rapid neutron-capture process (the $r$-process, which produces around half of the isotopes of the heaviest chemical elements; see \citet{Burbidge57,Cameron57,Frebel18,Cowan21}). For example, the surviving UFD Reticulum II contains highly $r$-process enhanced stars, implying it was enriched by a prolific early $r$-process event such as a neutron star merger \citep{Ji16,Ji16b,Roederer16}. Tucana III and Grus II also exhibit $r$-process enhancement \citep{Hansen17,Hansen20_Grus}. Satellite galaxies like these are located over 25 kpc away from the Sun \citep{Drlica15}, however, so studying their stars to learn about early chemical enrichment can be difficult.

Because the Milky Way was assembled hierarchically from many neighboring systems including UFDs, bona-fide dwarf galaxy stars can also be found located throughout our galaxy today, including near the Sun. Chemical tagging, i.e. using stellar chemical abundances to identify stars that formed together, is a promising way to identify dispersed UFD stars. Utilizing the Caterpillar simulation suite \citep{Griffen16} and a simple model for star formation and parametrized element enrichment, \citet{Brauer19} suggested that the population of galactic metal-poor $r$-process enhanced halo stars could have largely originated in UFDs. This idea stems from both observations of surviving UFDs such as Reticulum II, and kinematic studies of $r$-process stars \citep{Roederer18,Gudin21} that appear to be chemically and dynamically linked. Further evidence in support of chemically tagging $r$-process enhanced halo stars remains limited due to small sample size of known stars, but the $R$-Process Alliance \citep{Hansen18,Sakari18,Ezzeddine20,Holmbeck20} is continuing to discover more of these stars which should soon provide a rich sample for study. Low-mass galaxies, especially UFDs, also host a higher percentage of metal-poor stars compared to higher-mass galaxies \citep[e.g.,][]{Kirby13}. Chemical tagging with $r$-process elements and/or low-metallicity stars may thus help astronomers identify stars from UFDs.

Alongside chemical tagging, stellar dynamics also retain important information about the disrupted galaxies accreted by the Milky Way. In particular, the orbital actions and energy of a star are quasi-conserved quantities which can, in principle, be used to identify stars that were accreted together (see Section \ref{kinematics}). While these quantities are not truly conserved in the galaxy on long timescales, clustering in $E-L_z-J_r-J_z$ phase space (or a subset of this space) is a common, useful method to search for accreted structure. Thanks to the \textit{Gaia} mission, detailed 6D phase space information is now available for millions of stars \citep{Gaia18}. This influx of data has already lead to a better understanding of the major mergers that the Milky Way experienced (e.g., the \textit{Gaia} Sausage, \citealt{Belokurov18,Helmi18}, Sequoia, \citealt{Myeong19}, Kraken, \citealt{Kruijssen19,Kruijssen20,Forbes20}, and more, \citealt{Naidu20,Mardini22}). However, the low-mass galaxy mergers are far less understood because far fewer stars are contributed to the galaxy from each accreted UFD, rendering the associated dynamic signatures less pronounced and more difficult to isolate.

Currently, several groups are using kinematics to identify groups of stars that may have originated in UFDs. \citet{Roederer18} explored the possibility of identifying groups of stars that possibly originated together in UFDs by clustering stars with $r$-process enhancement (``$r$-process stars'') in dynamic phase space. \citet{Gudin21} expanded on this idea with a much larger data set of 446 stars. Both papers found multiple dynamically linked groups of stars, suggesting that these groups may represent dissolved UFD remnants and that dynamic clustering is indeed a promising method to identify groups of stars from tidally disrupted UFDs. Similarly, \citet{Limberg21} and \citet{Yuan20} used clustering algorithms to identify dynamically linked groups among very metal-poor ([Fe/H] $< - 2$) stars, several of which have similar dynamics to $r$-process enhanced stars.

This area of research is continuously expanding as more groups explore clustering with stellar dynamics -- both with and without chemical tagging -- as a means to identify possible groups of accreted stars from dwarf galaxies. And as astronomers continue to gather kinematics for millions of stars in our Galaxy, the search for these dwarf galaxy remnants is a difficult but worthwhile endeavor. It is unclear, however, to what degree we can trust the clusters identified by different clustering algorithms, and which clusters are most likely to correspond to real UFD remnants.

In this paper, we explore the possibilities and challenges of kinematically identifying stars from tidally disrupted UFDs in the Milky Way by analyzing a set of 32 cosmological zoom simulations of  Milky Way-mass galaxies. Using the \textit{Caterpillar} simulation suite \citep{Griffen16}, we trace tagged particles from accreted UFDs to $z=0$ and test different clustering algorithms in dynamic phase space. Specifically, we explore what fraction of remnant UFDs can be expected to be recovered using basic clustering algorithms, which clustering algorithms work best and most reliably, and which identified dynamically linked groups are most likely to correspond to real UFD remnants. In this work, we focus exclusively on UFDs because prior work has investigated more massive accretion events \citep[e.g.,][]{Wu22}, but UFDs remain poorly understood. While most cosmological simulations do not properly resolve UFDs, the \textit{Caterpillar} simulation suite provides us the unique ability to investigate many different Milky Way-mass galaxies forming in a cosmological context while resolving UFDs.

Section \ref{sec:sims} describes how we created simulated stellar halos from dark matter cosmological simulations, focusing on the methodology of tagging dark matter particles as tracers of stellar material and measuring the corresponding dynamics at $z=0$. Section \ref{sec:clus} describes seven different  clustering algorithms and how we test them on different data sets. Section \ref{sec:clusresults} discusses our clustering results and their implications for kinematically identifying UFD remnants in real data sets. Section \ref{sec:real} discusses the properties of real clusters and how to identify which clusters are most likely to correspond to real accreted UFD remnants. Section \ref{sec:takeaways} summarizes the takewaways for clustering observational data sets to best identify stars from accreted UFDs.

\section{Simulated Stellar Halos} \label{sec:sims}

\subsection{Cosmological Simulations}
We simulate stellar halos using 32 dark-matter-only cosmological simulations from the \textit{Caterpillar Project} \citep{Griffen16}. Each zoom-in simulation models the formation of a Milky Way-mass dark matter halo down to $z=0$. The effective resolution is $16,384^{3}$ particles of mass $3\times 10^4$ $M_\odot$ in and around the galaxies of interest, resolving subhalos down to total mass $\sim 10^6$ $M_\odot$. We limit our analysis to simulated Milky Way-mass halos that experienced no recent major merger; all other aspects of the accretion history are unbiased.

The simulations are fully described in \citet{Griffen16}. The halos in the zoom-in simulations were selected from a larger, lower resolution parent simulation with cosmological parameters from Planck 2013 $\Lambda$CDM cosmology: $\Omega_m = 0.32$, $\Omega_\Lambda = 0.68$, $\Omega_b = 0.05$, $\sigma_8 = 0.83$, $n_s = 0.96$, and H = 100 $h$ km s\textsuperscript{-1} Mpc\textsuperscript{-1} = 67.11 km s\textsuperscript{-1} Mpc\textsuperscript{-1} \citep{planck14}. Initial conditions were constructed using \texttt{MUSIC} \citep{music}.
Dark matter subhalos were identified using a modified version of \texttt{ROCKSTAR} \citep{rockstar,Griffen16} and merger trees were constructed by \texttt{CONSISTENT-TREES} \citep{consistenttrees}. The halos were assigned a virial mass $M_{vir}$ and radius $R_{vir}$ using the evolution of the virial relation from \citet{Bryan98}. For our cosmology, this corresponds to an overdensity of $\Delta_{crit} = 104$ at $z=0$.

\subsection{Dark Matter Particles as Tracers of Stellar Material}

Since the \textit{Caterpillar} simulations do not directly simulate stars, we tag dark matter (DM) particles as tracers of the stellar material of each accreted galaxy. Stars form tightly bound to their halos and move within the same potential as the dark matter, so a fraction of the most bound DM particles are expected to trace the phase-space distribution of the stars \cite[e.g.,][]{Bullock05,Cooper10}. We refer to the tagged particles as ``star particles'' and trace their phase-space distribution down to $z=0$.

There is debate over what fraction of DM particles should be tagged as tracers. The fraction generally ranges from the most bound 1-3\% \citep{Cooper10,Rashkov12,Bailin14}, to 5\% \citep{LeBret17,Cooper17,Dooley16}, to 10\% \citep{DeLucia08,Morinaga19,Tumlinson10,Gomez12}. \citet{Cooper17} finds that a fractions of 1-10\% all provide a good approximation to accreted halos of Milky Way analogs, implying that results for accreted galaxies are holistically insensitive to the exact fraction. Our analysis in this paper focuses on ultra-faint dwarf galaxies ($M_{halo} \leq 10^9 M_\odot$), so to ensure a sufficient number of particles to assess clustering, we tag the 5\% most bound particles. At this resolution, each tagged particle in an accreted ultra-faint dwarf galaxy corresponds to $\sim 10 M_\odot$ of stellar material. 
We note that having a single, fixed fraction is a simplifying assumption that breaks down in regions dominated by the baryonic potential and having significant angular momentum, such as the Milky Way disk \citep{Cooper17}. However, given that we focus on
dwarf galaxies in our analysis which are dark-matter dominated and elliptical, assuming a fixed fraction is not a principal concern. 

We tag the 5\% the most bound DM particles at the snapshot where the accreted halo reaches its peak mass. Alternative methods include tagging the particles at the snapshot before the halo is accreted or ``live'' tagging where stellar mass is added at each snapshot while the galaxy is star-forming. Our analysis focuses on small galaxies that are generally no longer forming stars at the time of their accretion, so we choose the peak mass as the snapshot at which to tag DM particles. We use a $M_{star}\sim M_{peak}$ relation to estimate the amount of stellar material represented by each tagged particle \citep{GarrisonKimmel17}. We note for completeness that live tagging would likely produce a more accurate phase space distribution but the significantly increased computational expense is beyond the scope of this work.

\begin{figure*}[tbp]

\gridline{
	\fig{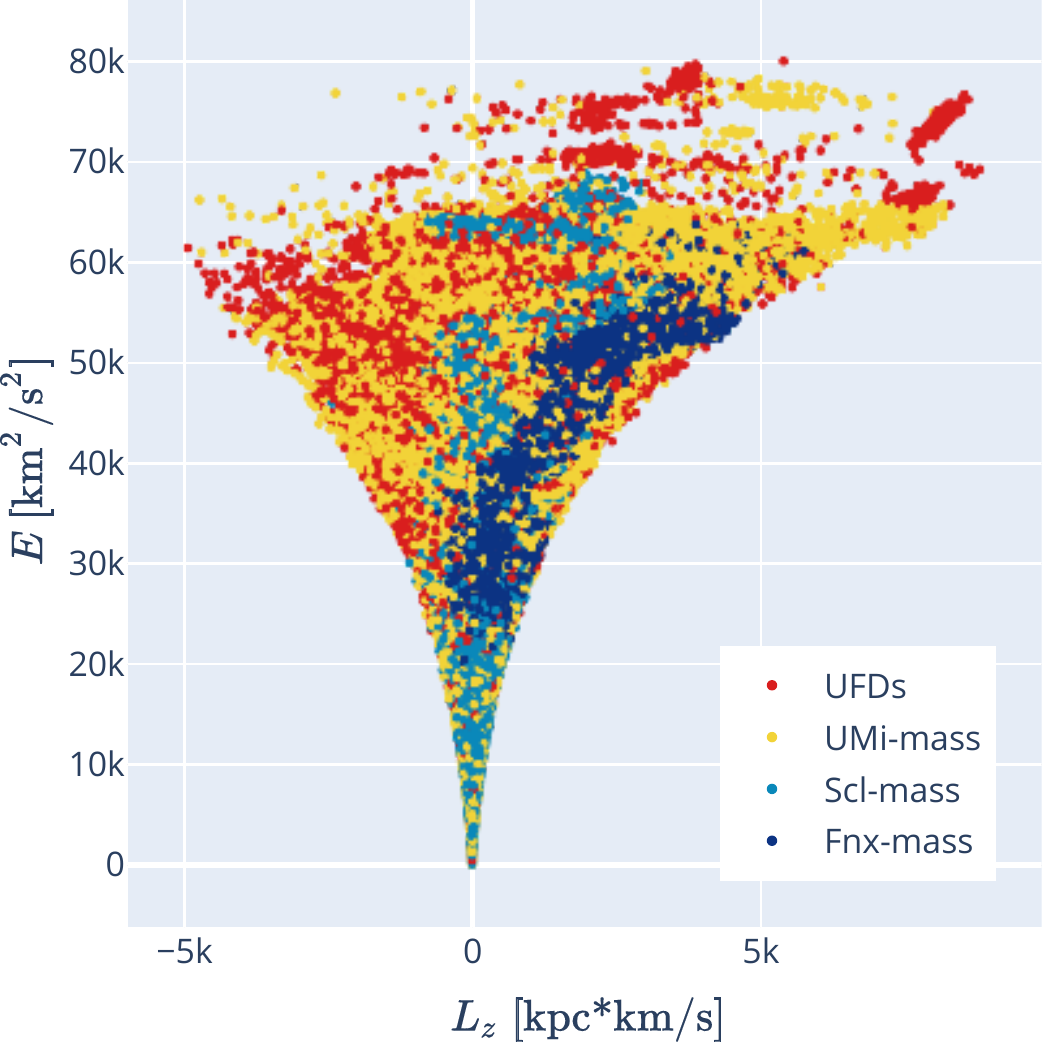}{0.35\textwidth}{}
	\fig{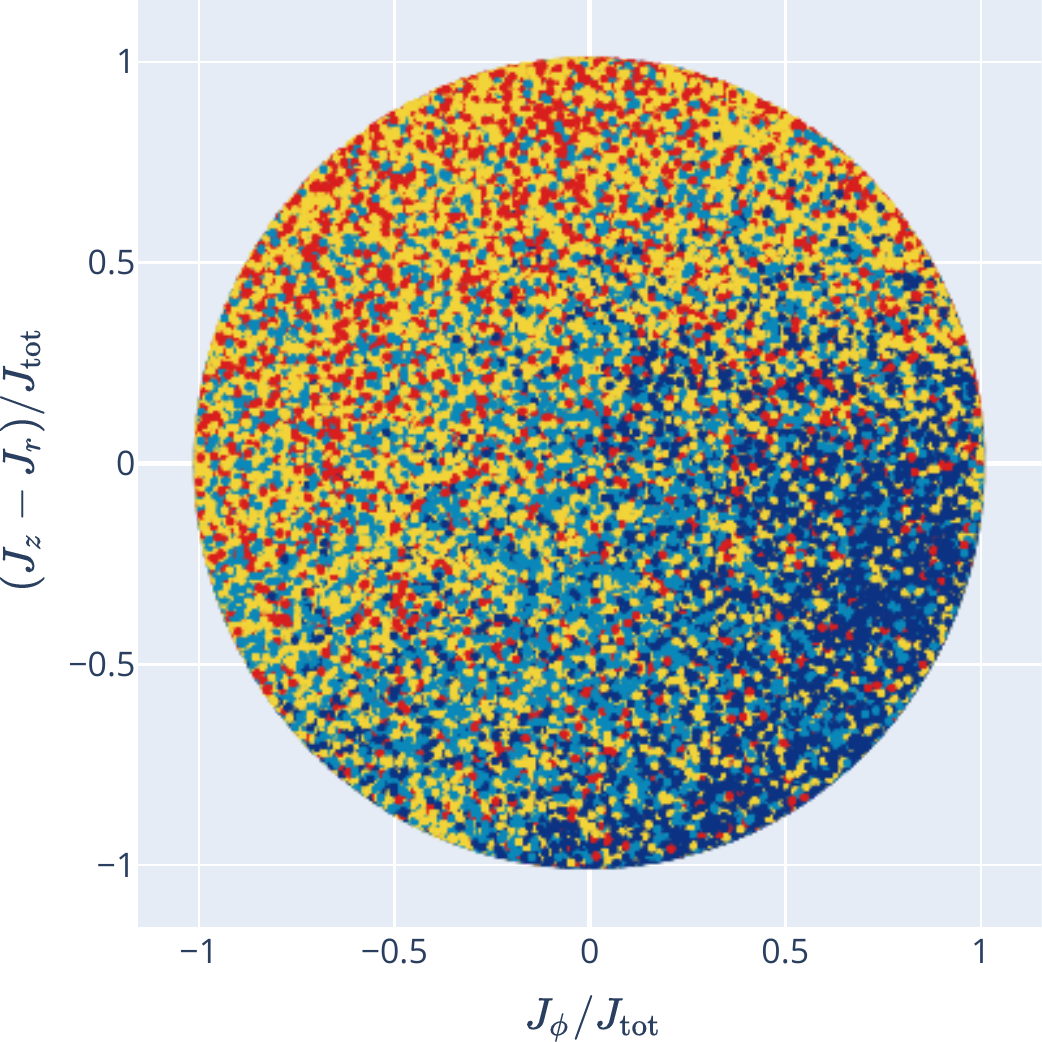}{0.35\textwidth}{}}
\caption{Left: $z=0$ dynamics (energy and $z$-angular momentum) of all accreted star particles within 50 kpc of the Sun in simulation Cat-14, one of the 32 simulated Milky Way-mass galaxies. The color of each particle corresponds to the mass of the galaxy in which it formed. Star particles from the smallest galaxies, the UFDs, are seen in red.
Right: $J_r$ and $J_z$ orbital actions for the same star particles.
\label{fig:allactions} }

\end{figure*}

While particle tagging is an imperfect method, it has repeatedly been shown to qualitatively capture trends and produce accreted stellar populations with properties (e.g., metallicities, spatial distribution, velocity dispersions) in agreement with observations around the Milky Way \citep[e.g.,][]{Cooper17,Rashkov12}. Given that this study is concerned with the qualitative situations in which kinematic clustering of accreted stars does or does not excel, particle tagging of dark-matter cosmological simulations is an ideal technique as a means to explore such clustering effects in our set of many different Milky Way-mass simulations. Moreover, a simulation with a disk would result in enhanced tidal disruption and phase space diffusion \citep{Errani17,Maffione18}, but because our results highlight the difficulty of identifying UFD remnants via clustering, our point is merely strengthened by our use of N-body simulations without an added disk potential.

\subsection{Stellar Dynamics} \label{kinematics}

We determine the dynamics of each accreted star particle (tagged DM particle) at $z=0$. In axisymmetric galactic potentials, stellar orbits are described by three integrals of motion called the orbital actions: $J_r$, $J_z$, and $J_\phi$ \cite[see][\S3.5]{Binney08}. Energy is another constant of motion for time-invariant potentials which, while not independent of the orbital actions, is useful during clustering searches. These four quantities are not conserved in realistic, time-varying galactic potentials, and the galactic potentials in the \textit{Caterpillar} simulations, for example, are approximately constant for only the last 5 Gyr or so ($z\lesssim0.5)$ \citep{Griffen16}. Despite this, these quantities provide a useful phase space in which to search for dynamically-similar stars that is currently being used by several groups in the search for stars from UFDs. We thus explore the possibilities of using these dynamics. These integrals of motion are defined as \citep{Binney12}:
\begin{enumerate}
    \item $E$: the specific orbital energy, the total orbital energy of the star divided by its mass.
    \item $J_r$: the orbital action that quantifies oscillations of an orbit along the radial direction. $J_r$ is non-negative and increases for more eccentric orbits.
    \item $J_z$: the orbital action that quantifies oscillations about the equatorial plane. $J_z$ is non-negative and increases for orbits that rise more out of the equatorial plane.
    \item $J_\phi$: the azimuthal orbital action, equal to the angular momentum out of the equatorial plane ($J_\phi = L_z$). 
\end{enumerate}

To estimate orbital actions, one first needs an initial estimate of the gravitational potential. For each of our 32 simulations, we use the AGAMA software library \citep{Vasiliev19} to construct an estimated axisymmetric gravitational potential. The potential is built via multipole expansion in spherical harmonics with $l_{max}=8$, using the locations and masses of all N-body particles at $z=0$. We validate the estimated potential by comparing it to the value of the potential stored for each particle from the original \textit{Caterpillar} simulation, confirming the same relative potential energy between particles. After constructing the axisymmetric potential, we use the galactocentric positions and velocities of each accreted star particle to compute the associated actions within AGAMA.

As an illustrative example, the $z=0$ phase space distribution for the accreted star particles in one of our simulations can be seen in Figure \ref{fig:allactions}. The particles in these plots are colored based on the peak mass of the galaxy in which each of them formed: UFD ($M_*\leq10^{5}M_\odot$), Ursa Minor-mass ($M_*=10^{5} \text{ to } 10^{6} M_\odot$), Sculptor-mass ($M_*=10^{6} \text{ to } 10^{7} M_\odot$), and Fornax-mass ($M_*=10^{7} \text{ to } 10^{8} M_\odot$). Note that this example galaxy did not accrete more massive dwarfs such as those with masses similar to that of the Large Magellanic Cloud. 

In Figure \ref{fig:allactions}, the particles from UFDs are only 9\% of all the accreted particles within this radial cut, but they are still identifiable in the outskirts of the phase space diagram because virtually all of the particles from more massive dwarfs are overlap significantly in phase space. This implies we may be able to more easily identify some UFD remnants at, for example, high energy.

\begin{figure}[tb]
\center
\includegraphics[width=0.9\linewidth]{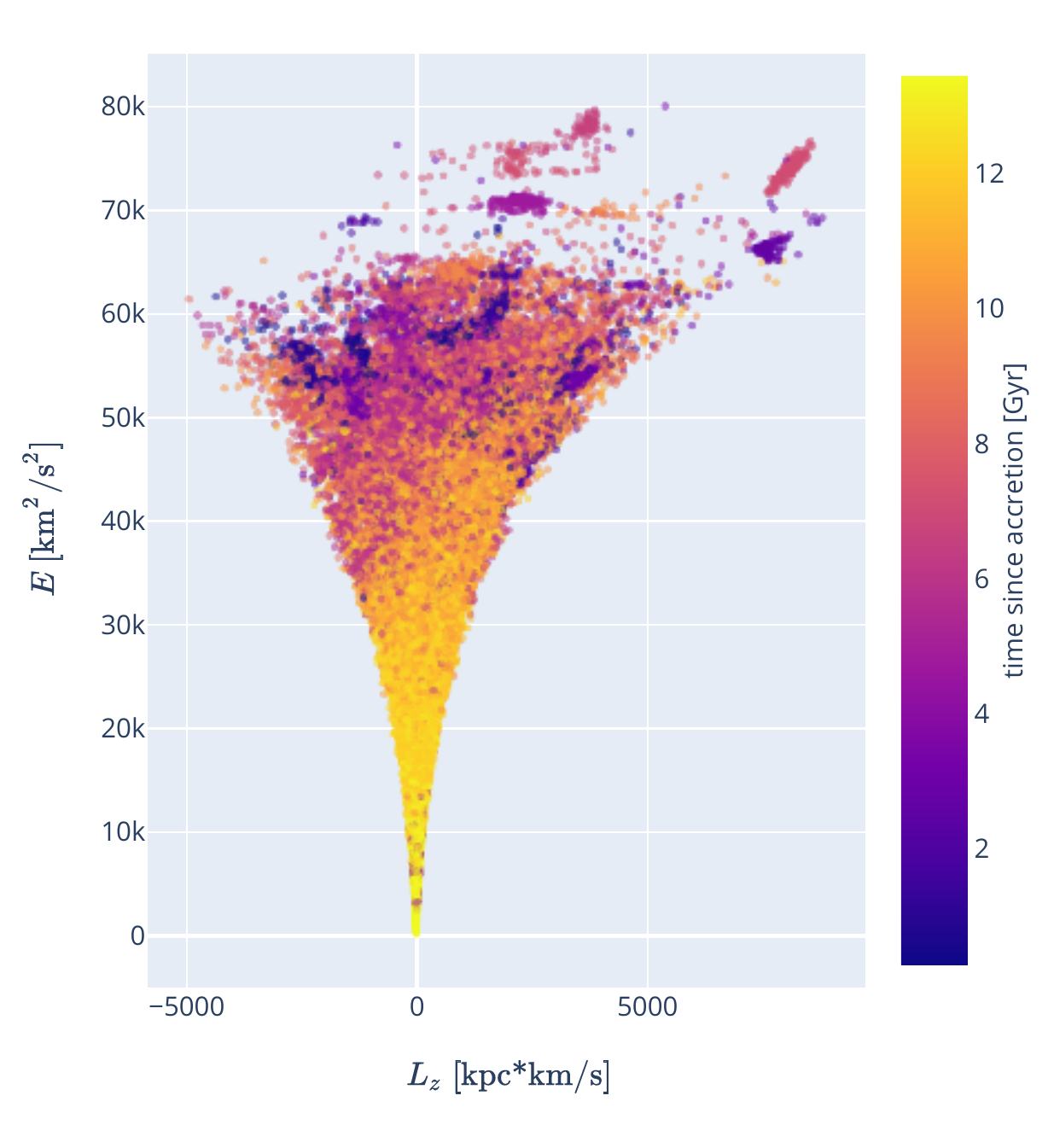}
\caption{
$z=0$ dynamics of star particles that originated from UFDs in one of our simulations (with a 50 kpc radial cut; see Section \ref{subsec:datsets}). 
The color of each particle corresponds to how long ago it was accreted by the Milky Way-mass host galaxy. Stars that were accreted more recently are, generally, of higher energy and less phase mixed. Over time, the stars mix more in the phase space and are less identifiable by clustering algorithms.
\label{fig:infall}}
\end{figure}

Specifically considering the particles from UFDs, in Figure \ref{fig:infall}, we show that any identifiable remnants are from relatively recent accretion events, while the most phase-mixed particles are from accretion events that occurred over 8 Gyr ago. This is to be expected, since more recent accretion events will have maintained a stronger dynamic signature at $z=0$ compared to stars that have been relaxing in the stellar halo for many gigayears \citep[e.g.,][]{Gomez10}.

\subsection{The Different Data Sets We Consider} \label{subsec:datsets}

We consider how well clustering works for data sets with three different radial cuts at varying distances from the Sun: 
\begin{enumerate}
    \item All accreted star particles, no radial cut. This is a complete data set which cannot be produced with real observations.
    \item All accreted star particles within 50 kpc of the Sun. This is an idealistic data set that extends to roughly where the stellar halo drops off.
    \item All accreted star particles within 5 kpc of the Sun. This is a more realistic data set that includes stars for which we can obtain decent parallax measurements from Gaia.
\end{enumerate}

The location of the ``Sun'' in each simulation is a consistent, randomly chosen location in the equatorial plane 8 kpc from the galactic center.

We also consider data sets with:
\begin{enumerate}
    \item Only accreted star particles from UFDs. This data set is idealistic. To pursue it observationally, one could focus on limiting to only stars with certain chemical signatures (e.g., low metallicity, $r$-process enhancement, deficiency in neutron-capture element abundances) and/or removing stars that are known to be associated with larger mergers.
    \item All accreted star particles.
\end{enumerate}

After matching each radial cut with UFD-only and all-stars data sets, we have a total of six data sets. Each data set includes stellar dynamics from 32 simulations (though not all simulations are used when performing clustering analysis of the larger radial cuts due to computational limitations). We then quantify how well each clustering algorithm performs in these six situations.

\begin{figure}[tb]
\center
\includegraphics[width=0.9\linewidth]{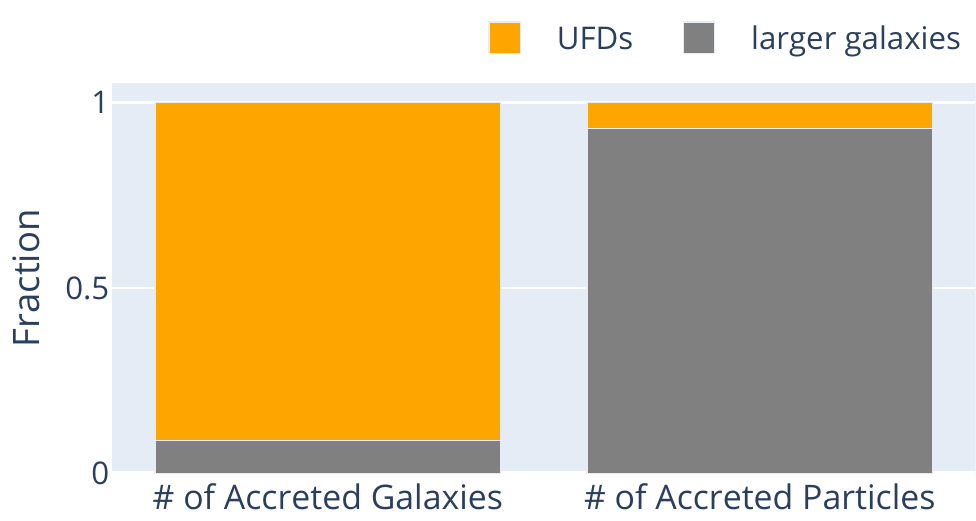}
\caption{
Across all 32 simulations, $\sim91\%$ of the galaxies accreted by Milky Way-mass galaxies are ultra-faint dwarf galaxies. These small galaxies only contribute $\sim 7\%$ of the accreted star particles, however. These fractions are roughly constant with radial cut.
\label{fig:UFDfrac}}
\end{figure}

For the data set without a radial cut, Milky Way-mass galaxies accrete on average $187^{+69}_{-65}$ UFDs. This is $91^{+1}_{-1}\%$ of the total number of accreted systems that Milky Way-mass galaxies will ever accrete. Despite UFDs being the vast majority of accreted galaxies, though, they only contribute $\sim 7\%$ of the accreted star particles. These fractions are shown in Figure \ref{fig:UFDfrac}. These results align with \citet{Monachesi19}, which estimated that the accreted stellar halo had only a handful of significant progenitors. For the data set with a 5 kpc radial cut, the total average number of accreted UFDs seen in the data set drops to $99^{+45}_{-30}$ but the percentage representation remains the same. We note here that all uncertainty values provided represent 16th -- 84th percentile scatter across all the simulations.

\section{Clustering Methodology} \label{sec:clus}

\subsection{Clustering Algorithms} \label{subsec:algos}

We apply seven different clustering algorithms on the four-dimensional energy-action space of each simulated Milky Way-like halo. The algorithms studied in this work are HDBSCAN \citep{Campello15,McInnes17}, Gaussian mixture models \citep[GMM;][]{Dempster77}, agglomerative clustering \citep{Ward63}, K-means \citep{Lloyd82,Vassilvitskii06}, affinity propagation \citep{Frey07}, mean-shift \citep{Comaniciu02,Derpanis05}, and friends-of-friends \citep{Huchra82,Press82,Davis85,pyfof}. Before running any clustering alogithms on our simulations, we normalize each of the 4D energy-action variables into the range [0, 1]. Here, we briefly comment on each of these algorithms.

HDBSCAN (Hierarchical DBSCAN) is a hierarchical extension of the density-based approach of DBSCAN. It measures the density around each point, constructs a hierarchical cluster tree based on this density information, and returns clusters that are persistent across different density thresholds. As a result, it is sensitive to datasets having true groups at varying densities. It also scales well for massive datasets. \citet{Hunt21} found that, compared to DBSCAN and GMM, it performs best at recovering open clusters in a massive sample of Gaia data. This was also the preferred clustering algorithm of \citet{Gudin21} and \citet{Limberg21}, two papers that identified dynamically linked groups that may correspond to UFDs. 


Agglomerative clustering forms clusters from the bottom up. It starts with each particle as its own cluster. Clusters that are separated by the least linkage distance (in our case, Euclidean distance) are then hierarchically merged until the pre-set number of clusters is reached. Because it has a time complexity of $O(n^3)$ and requires $\Omega(n^2)$ of memory, it is too slow and memory-intensive for large datasets. 

K-means is a distance-based algorithm that returns a pre-set number of k clusters, each of equal variance. Starting with k randomly generated initial means, it first assigns each particle to the mean with the least sum-of-squares distance. Particles associated with the same mean form a cluster. The mean (or centroid) of each cluster—and consequently, cluster membership—is then continually updated until convergence.

A Gaussian mixture model can be thought of as a generalization of K-means in that it returns distance-based clusters which may be at different variances. It decomposes the sample into a mixture of a pre-set number of n Gaussian distributions and upon convergence, returns the Gaussian components as separate clusters. 

Unlike K-means, agglomerative clustering, and Gaussian mixture models, affinity propagation does not require a pre-set number of clusters before running. Its goal is to find “exemplars” or prototype particles that are representative of a cluster. First, each particle begins as a potential exemplar. Pairs of particles then pass “messages” to each other about suitability of one particle to be the exemplar of the other. These messages are passed until a stable set of exemplars and, thus, clusters emerge.

Mean-shift is a centroid-based algorithm that treats each particle as a kernel with a pre-set bandwidth. It then performs a gradient ascent on the kernel peaks until convergence. \citet{Gomez10} used mean-shift on the $E-L-L_z$ space of a mock Gaia catalogue of the solar neighborhood and recovered roughly 50\% of all satellite galaxies. We note that this differs from our results because this work focused on a smaller quantity of larger-mass satellites as compared to our UFD-focused analysis.

Friends-of-friends (FoF) is commonly used to identify gravitationally bound halos in cosmological simulations. Particles that are separated by a distance less than a pre-set linking length are linked as “friends,” forming a networked cluster of particles. Networks that have no mutual friends are designated as separate clusters. \citet{Helmi00} applied this algorithm on the $E-L-L_z$ space of a mock Gaia catalog to identify simulated Milky Way accretion events.

Other groups have used custom clustering algorithms, e.g. StarGo \citep{Yuan18,Yuan20}, Enlink \citep{Sharma09,Wu22}, and other hierarchical clustering techniques \citep{Lovdal22,RuizLara22}. We do not test all of these algorithms, but expect our UFD-focused results to holistically hold for them as well (see Section \ref{subsec:struggle}).


\subsection{Hyperparameter Choices} \label{subsec:hyperparam}

All the algorithms included in this paper except affinity propagation require a pre-selected hyperparameter in order to begin clustering. To explore different hyperparameter choices, for each algorithm we:

\begin{enumerate}
    \item Create a hyperparameter search space consisting of about 20 trial values. For instance, to select the \texttt{min\_cluster\_size} hyperparameter for HDBSCAN, we create a search space composed of integers from 3 to 20 inclusive, and for FoF we explore from 0.001 to 0.2.
    \item Run the clustering algorithm with each trial hyperparameter on each simulation in each data set.
    \item For every clustering run, count the number of pure and complete clusters. A cluster is ``pure'' if $\geq \frac{2}{3}$ of the stars in that cluster accreted together from a UFD. A cluster is also ``complete'' if $\geq \frac{1}{2}$ of the stars from that accreted UFD are found together in that cluster.
    
    
    \item For every simulation on which a particular hyperparameter is tested, calculate a recovery rate and a realness rate. The recovery rate is defined as: \[\frac{\text{number of pure and complete clusters}}{\text{number of accreted UFDs in the data set}}\times100\%\]
    
    Meanwhile, the realness rate is defined as: \[\frac{\text{number of pure clusters}}{\text{number of clusters found by the algorithm}}\times100\%\]
    
    When calculating these rates, we only consider clusters and remnants with at least 5 particles.
    
    
    \item For each data set, determine the optimal hyperparameter by assigning a score to each hyperparameter choice. To assign the score, normalize all of the recovery rates and realness rates using a min max scaler, and then add the normalized median recovery and realness rates together. The optimal hyperparameter thus balances the highest UFD recovery rate and the highest realness of its clusters.
    
\end{enumerate}

We choose an optimal hyperparameter value for each algorithm on each data set. Since we are testing six algorithms that each require hyperparameters on six different data sets, we make a total of 36 optimized hyperparameter selections. A summary of the optimal hyperparameter choices are in the Appendix.

\subsection{Association of Different Observables with the True Cluster Labels} \label{subsec:anova}

\begin{figure}[t!]
\center
\includegraphics[width=\linewidth]{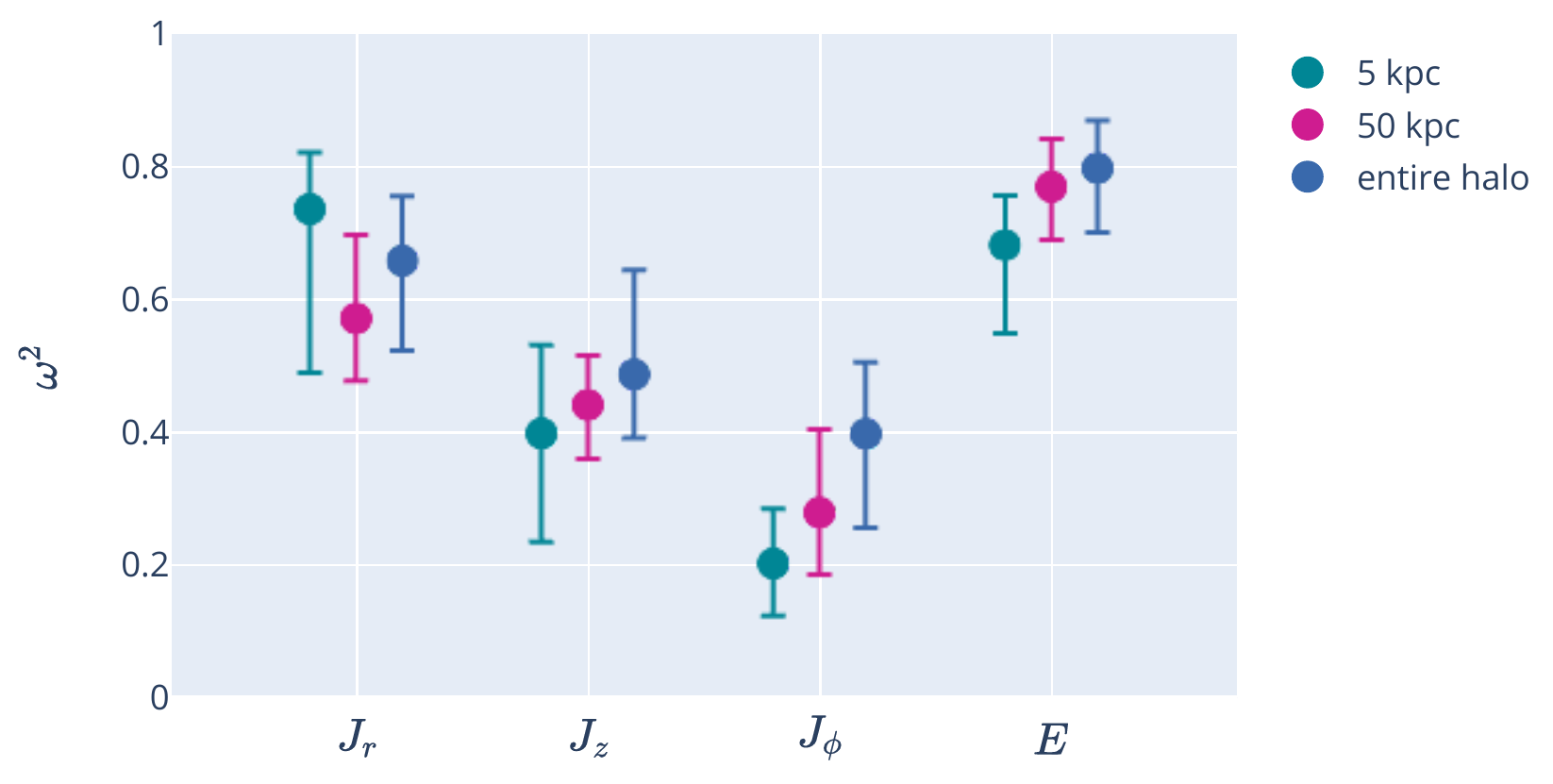}
\caption{Strength of association between actions ($J_r$, $J_z$, $J_\phi$, $E$) and the true cluster labels. Higher $\omega^2$ values indicate a stronger association. $E$ has the highest $\omega^2$ values, implying it is the most important variable when seeking to find clustered stars that accreted together. Note that $J_\phi$ here is equivalent to $L_z$.
\label{fig:omegasq}}
\end{figure}

\begin{figure*}[tb]
\center
\includegraphics[width=0.7\linewidth]{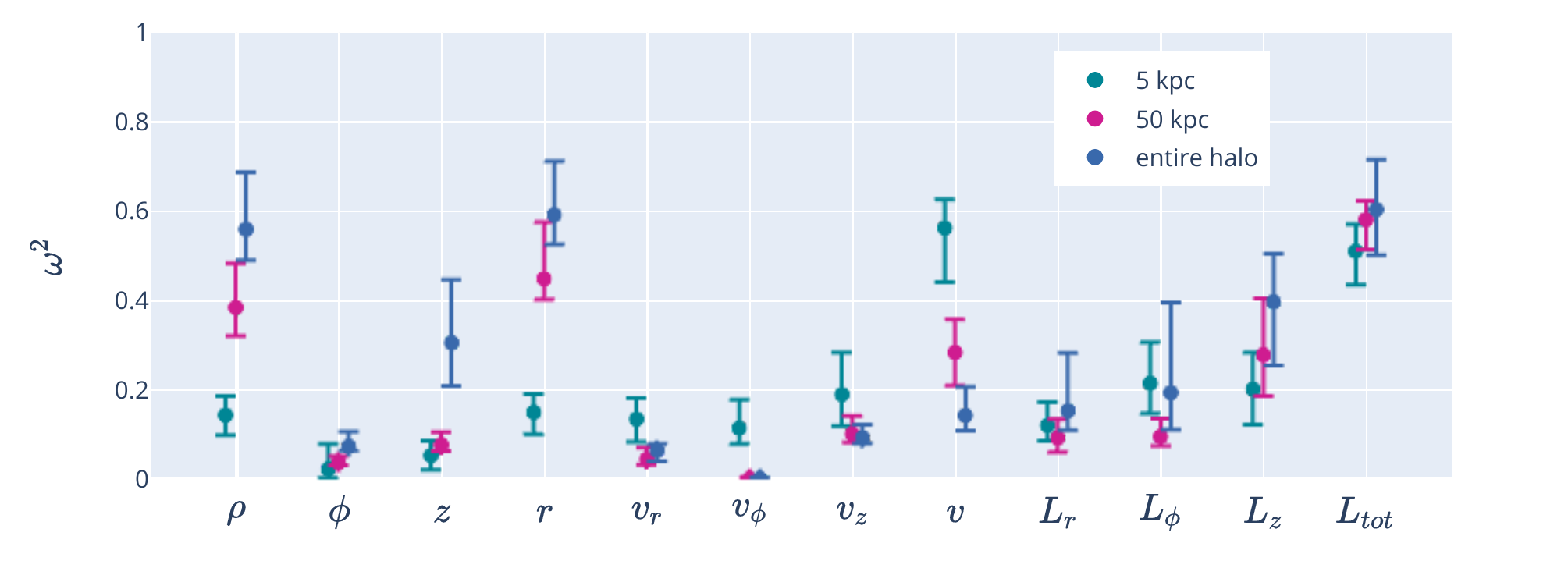}
\caption{Strength of association between different kinematic observables and the true cluster labels. Higher $\omega^2$ values indicate a stronger association. $L_{tot}$ has consistently high $\omega^2$ values, implying it is an important variable when seeking to find clustered stars that accreted together. $\rho$, $\phi$, $z$ and $v_r$, $v_\phi$, $v_z$ are the radius and velocity in cylindrical coordinates, respectively. The importance of $r$ and $v$ is due to their correlation with total energy.
\label{fig:omegasq2}}
\end{figure*}

To help identify which observable variables are most likely to be important during clustering, we perform one-way analysis of variance (ANOVA) tests on the stellar kinematics of each simulation. The ANOVA test assesses the association between a categorical (e.g., the label of each true cluster) and a continuous variable (e.g., each of the kinematic variables) \cite[e.g.,][]{mcdonald_2014,Gomez14}. If a given kinematic variable is strongly associated with the true cluster labels, it is likely to be important during clustering in situations where we do not know the true labels.

We use the \texttt{stats.f\_oneway} ANOVA test from the \texttt{scipy} python package \citep{2020SciPy-NMeth}. This F-test analyzes whether the means of the continuous variable differs between groups. $F = $ (variation between cluster means) / (variation within the clusters), so high $F$ values for our data signify that a given observable varies more between clusters than within. For these tests, the clusters we are using are the true UFD remnant groups because we take the labels directly from the simulations. To quantify the level of the effect, we also calculate the $\omega^2$ value of each test \cite[e.g.,][]{olejnik03}. This metric is similar to $R^2$ in the context of regression analysis while also accounting for the degrees of freedom in the model. $\omega^2$ can vary from $-1$ to $+1$; values far from zero imply a stronger effect.

The ANOVA test results are shown visually in Figures \ref{fig:omegasq} and \ref{fig:omegasq2}. Figure \ref{fig:omegasq} shows the four axisymmetric actions we use in clustering. All four actions show correlation with the true cluster labels, with energy consistently being the most important observable. Figure \ref{fig:omegasq2} shows the correlations of other potentially useful observables, demonstrating the high correlation of total angular momentum, $L_{tot}$. These results support our choice to cluster in $E$-$J_r$-$J_z$-$J_\phi$ phase space. They also imply that $E$-$L_{tot}$ phase space can be useful to find UFD remnants in cases where the full axisymmetric actions are unknown. This has been known previously \citep[e.g.,][]{Helmi00,Gomez10}.

Figure \ref{fig:omegasq2} shows that total velocity is likely important at parallax-level cuts (e.g., 5 kpc) and total distance from the galaxy's center is important for data sets with no radial cut. This is simply due to the relationship between velocity, radius, and total energy. All of the test results are summarized in the Appendix in Table \ref{tab:ANOVA}. As an additional check, we also include ANOVA tests for $z_{infall}$, the redshift at which the particles were accreted by the Milky Way. This variable perfectly aligns with the true cluster labels and thus should have $\omega^2=1$, which we find.

\section{Quantifying the Abilities and Limitations of Clustering Algorithms} \label{sec:clusresults}

We run each clustering algorithms (HDBSCAN, Gaussian Mixture Models, Agglomerative Clustering, Mean-Shift Clustering, K-Means, Friends-of-Friends, and Affinity Propagation; see Section \ref{subsec:algos}) on each simulation in each of the six data sets (see Section \ref{subsec:datsets}). The hyperparameters of each algorithm are chosen as described in Section \ref{subsec:hyperparam}. All clustering is done in 4D energy-action space ($E$, $L_z$, $J_r$, and $J_z$) as supported by the association results presented in Section \ref{subsec:anova}. Given the seven algorithms, six data sets, up to 32 simulations per data set, and roughly twenty hyperparameter choices per algorithm, we run over 10,000 clustering tests.

The results from these tests are largely a cautionary tale. All of these algorithms have significant limitations when it comes to identifying UFD remnant groups. Hence, in this section, we analyze the possibilities and limitations of the algorithms with a focus on how the results can inform the search for UFD remnants in real data sets since there currently exist no better methods to identify tidally disrupted ultra-faint dwarf galaxies from survey data. In future work, fully modeling the phase-space distribution of all accreted systems simultaneously could offer an alternative method to learn about accreted UFDs as compared to the current method of individually picking out a handful of dynamic clusters that may or may not correspond to UFDs. For now, though, kinematic clustering is one of the few available methods.

The basic problem is that, due to phase mixing and background, most star particles that accreted into the Milky Way-mass galaxies from the small UFD remnants overlap too much with other particles in phase space at $z=0$ to be reliably identified as coherent remnant groups. This is true for all algorithms across all data sets. The clustering algorithms also frequently return clusters that do not correspond to any true UFD remnant (``false positives''). However, some algorithms work better than others and some identified clusters are more likely to be real than others. We now give more details on algorithm usability.


\subsection{Example Clustering Results}


\begin{figure*}[p]
\center
\includegraphics[width=1\linewidth]{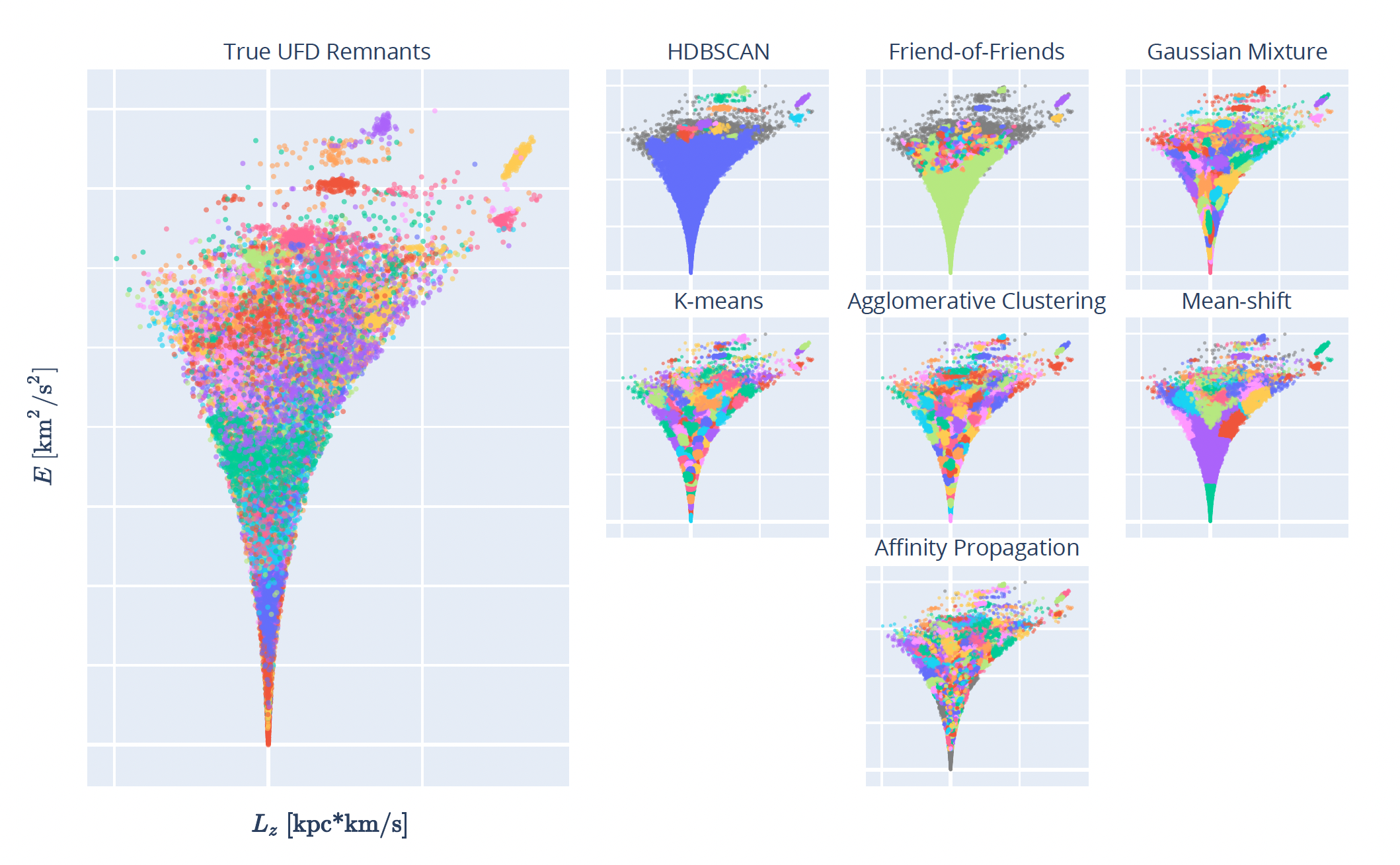}
\caption{Example of clustering results for one simulation (Cat-14) from one data set (accreted star particles from UFDs within 50 kpc of the Sun). Far left: Star particles from true UFD remnants in dynamic phase space. Many of the particles are phase mixed. Right: Results from each of the seven clustering algorithms tested in this paper. Most of the clusters found by these algorithms, especially those at lower energy, do not correspond to true UFD remnants.
\label{fig:algos}}
\end{figure*}

\begin{table*}[p]
\begin{tabular}{lll}
 \hline
Algorithm               & Realness Rate           & Recovery Rate                     \\ \hline
HDBSCAN                 & 67\% (12 pure clusters / 18 total clusters) & 4\% (5 pure \& complete clusters) \\
Friend-of-Friends       &   34\% (61 pure clusters / 176 total clusters)              &   5\% (6 pure \& complete clusters)         \\
Gaussian Mixture Models &      18\% (29 pure clusters / 160 total clusters)        &      5\% (6 pure \& complete clusters) \\
K-Means & 12\% (27 pure clusters / 230 real clusters) & 5\% (6 pure \& complete clusters) \\
Agglomerative Clustering & 13\% (32 pure clusters / 248 total clusters) & 6\% (8 pure \& complete clusters) \\
Mean-Shift & 22\% (24 pure clusters / 100 total clusters) & 3\% (4 pure \& complete clusters) \\
Affinity Propagation & 5\% (52 pure clusters / 989 total clusters) & 2\% (3 pure \& complete clusters)
\\  \hline
\end{tabular}
\caption{For the example simulation shown in Figure \ref{fig:algos}, the realness and recovery rates of different clustering algorithms. The recovery rate is determined by comparing the number of pure \& complete clusters to the total number of accreted UFDs in this simulation, 124 UFDs. \label{tab:example}}
\end{table*}

\begin{figure*}[tb]
\center
\includegraphics[width=0.85\linewidth]{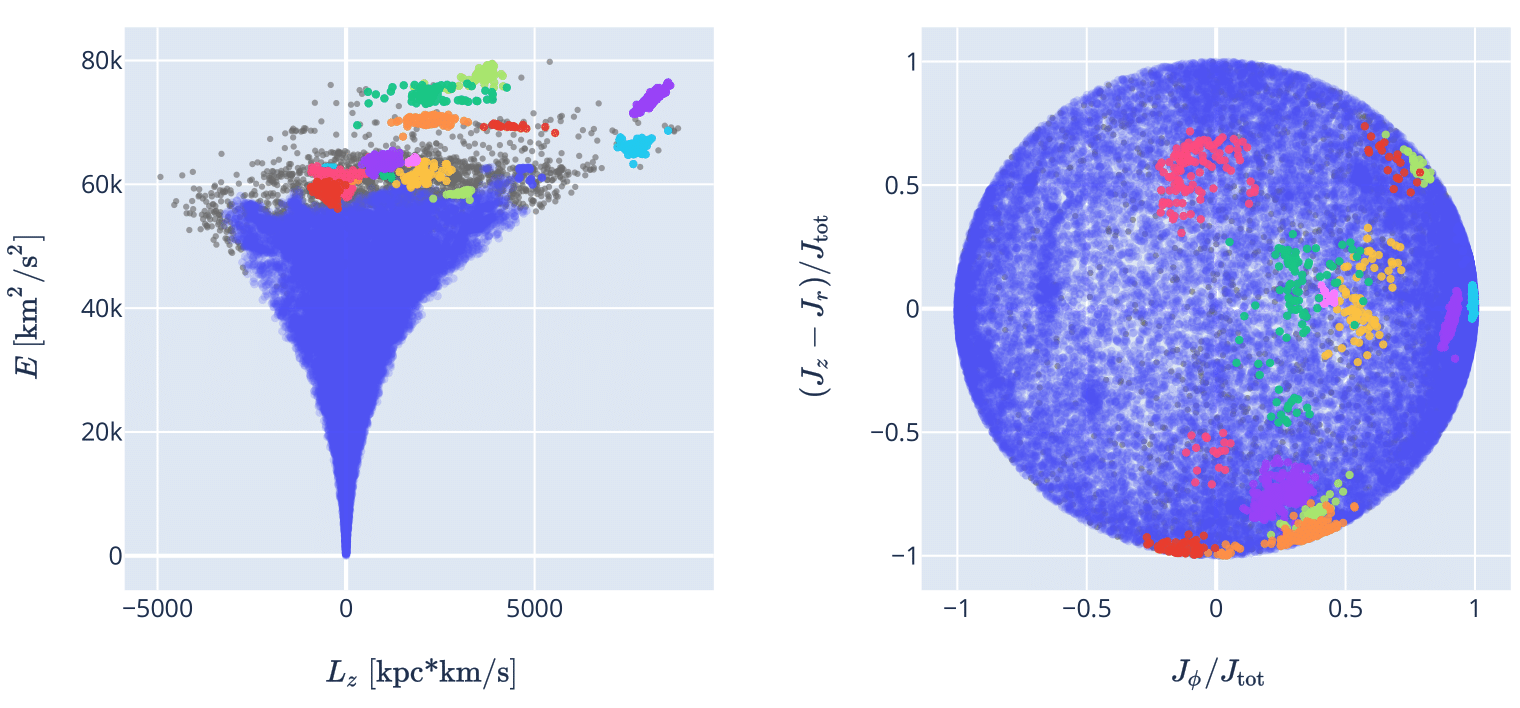}
\caption{
Example of the HDBSCAN clustering results for one simulation (Cat-14) from one data set (accreted star particles from UFDs within 50 kpc of the Sun). The clustering is done in 4D energy-action space ($E$, $L_z$, $J_r$, $J_z$). Grey points are particles not associated with any cluster. For this simulation, HDBSCAN finds eighteen clusters. Twelve of them are real groups of accreted UFD stars, and five of those twelve are fully ``recovered'' UFD remnants. 
\label{fig:HDBSCAN}}
\end{figure*}


Figure \ref{fig:algos} shows example clustering results from each of the seven algorithms. These results use a single Milky Way-mass simulation (simulation Cat-14) from one data set (accreted star particles from UFDs within 50 kpc of the Sun). The left shows the true UFD remnants in phase space; each star particle is colored according to the UFD it was born in (note that each color repeats several times). The star particles in this example originated in 124 different UFDs. The panels on the right show how well each clustering algorithm performs. All clustering algorithms perform poorly in the high density region of phase space and only consistently identify several isolated, high-energy clusters. These high-energy clusters do, in fact, correspond to real UFD remnants. The majority of the rest of the clusters found by these algorithms do not actually correspond to real UFD remnants. This is unsurprising given the high density of overlapping structure in the high density region.

For all of our clustering results, we use the metrics of ``realness rate'' and ``recovery rate'' to evaluate the findings. Realness rate is defined as the fraction of clusters which are ``pure'', defined as clusters for which at least $2/3$ of the stars accreted together. Recovery rate is defined as the fraction of UFD remnants which are recovered. A remnant is recovered if (1) its stars are clustered into a pure cluster and (2) that cluster is ``complete'', defined as clusters for which at least $1/2$ of the stars from a remnant are identified together in a single cluster. When determining these rates, we only consider clusters or remnants with at least 5 particles. The purity and completeness thresholds (2/3 and 1/2, respectively) are chosen with a stricter requirement on the ``realness'' of a cluster as our priority is identifying stars that accreted together. These thresholds can both be varied, though, and are simply chosen as example metrics. The holistic takeaways of this paper remain consistent even if you vary these thresholds.

As an illustrative example, the realness and recovery rates for each algorithm on the Cat-14 simulation are reported in Table \ref{tab:example}. The example HDBSCAN results are shown in Figure~\ref{fig:HDBSCAN}. 


\begin{figure*}[h]
\center
\subfloat{
\includegraphics[width=0.9\textwidth]{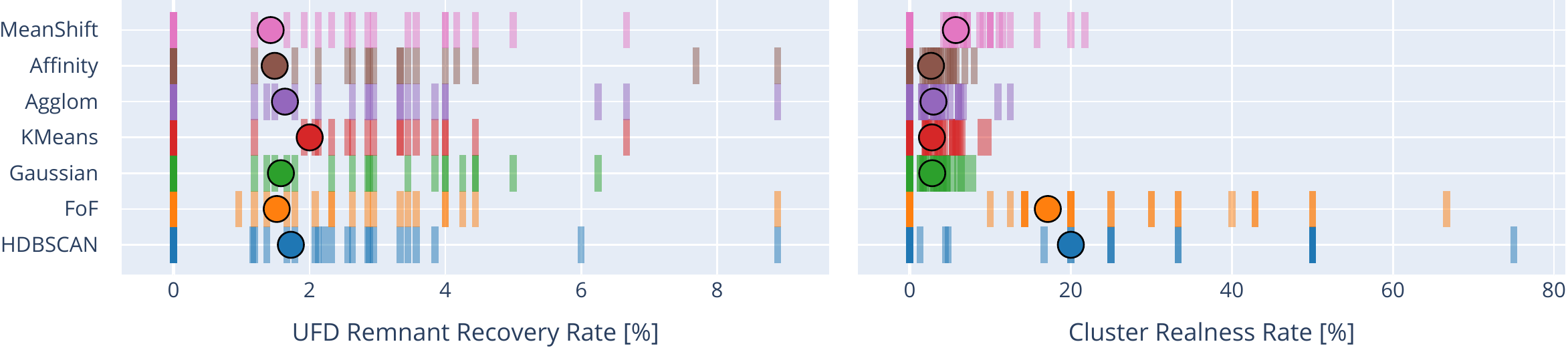}%
}

(a) UFD + 5 kpc data sets: All algorithms recover similarly low numbers of UFD remnants. HDBSCAN and FoF have the highest cluster realness rates. \\[4ex]

\subfloat{
\includegraphics[width=0.9\textwidth]{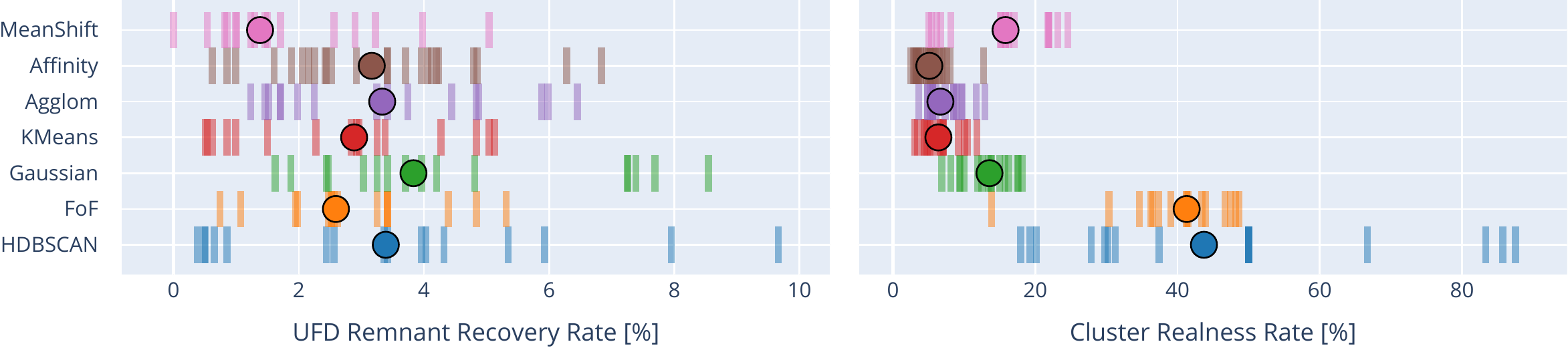}%
}

(b) UFD + 50 kpc data sets: Once again, HDBSCAN and FoF have the best balance of UFD remnant recovery and cluster realness rates. \\[4ex]

\subfloat{
\includegraphics[width=0.9\textwidth]{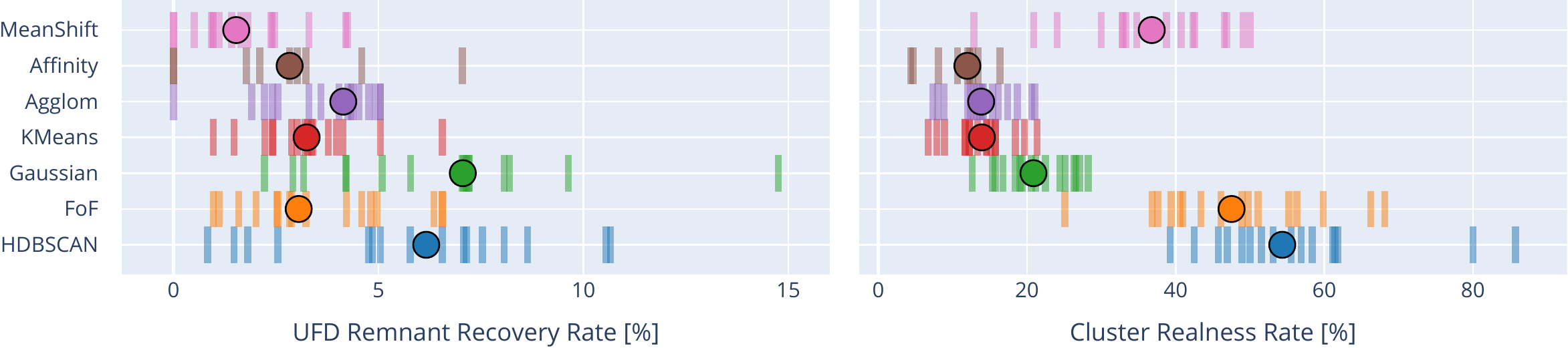}%
}

(c) UFD + entire halo data sets: HDBSCAN has the best balance of UFD recovery and cluster realness. FoF has a similar realness rate but recovers far fewer remnants. \\[4ex]

\caption{Results for the UFD-only data sets. Each line represents a single Milky Way-mass galaxy simulation and each circle is the median rate across all simulations. See Section \ref{subsec:datsets} for descriptions of the different data sets and Section \ref{subsec:hyperparam} for definitions of recovery and realness rates. Generally, HDBSCAN and FoF perform better than the other algorithms and are also significantly faster. For results with the all-stars data sets, see Figure \ref{fig:allalgos}. 
\label{fig:UFDalgos}}

\end{figure*}

\begin{figure*}[h]
\center
\subfloat{
\includegraphics[width=0.9\textwidth]{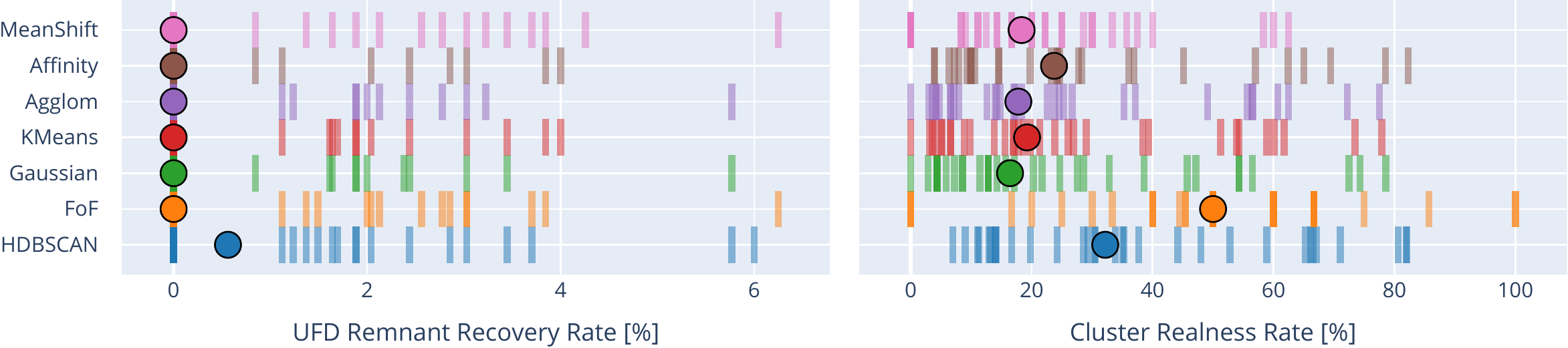}%
}

(a) All stars + 5 kpc data sets: Algorithms find real clusters accreted from dwarf galaxies, but almost none of them are UFD remnants. \\[4ex]

\subfloat{
\includegraphics[width=0.9\textwidth]{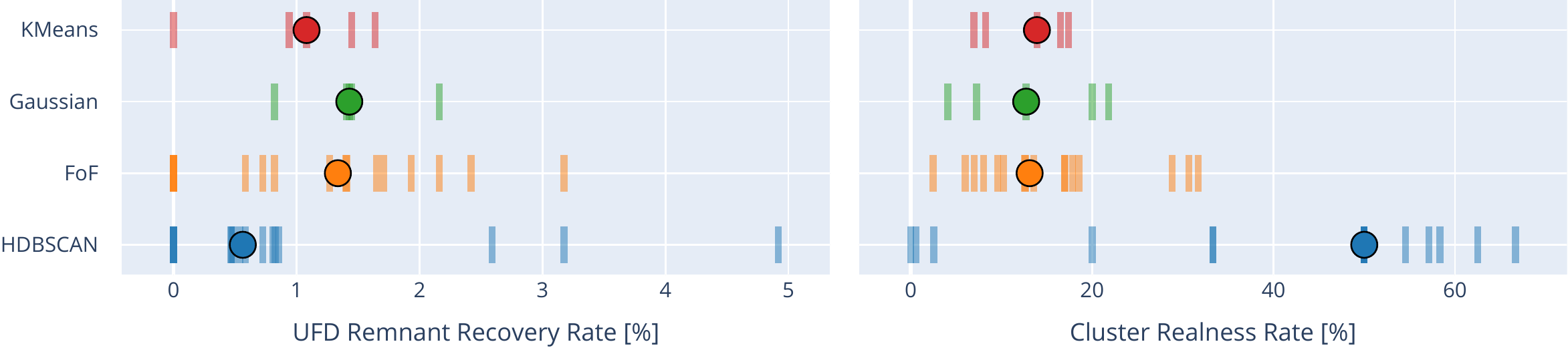}%
}

(b) All stars + 50 kpc data sets: Recovery rates are once again low, but realness rates can be high as clusters from larger mass dwarfs are identified. HDBSCAN has highest realness rate, but all recovery rates are low.\\[4ex]

\subfloat{
\includegraphics[width=0.9\textwidth]{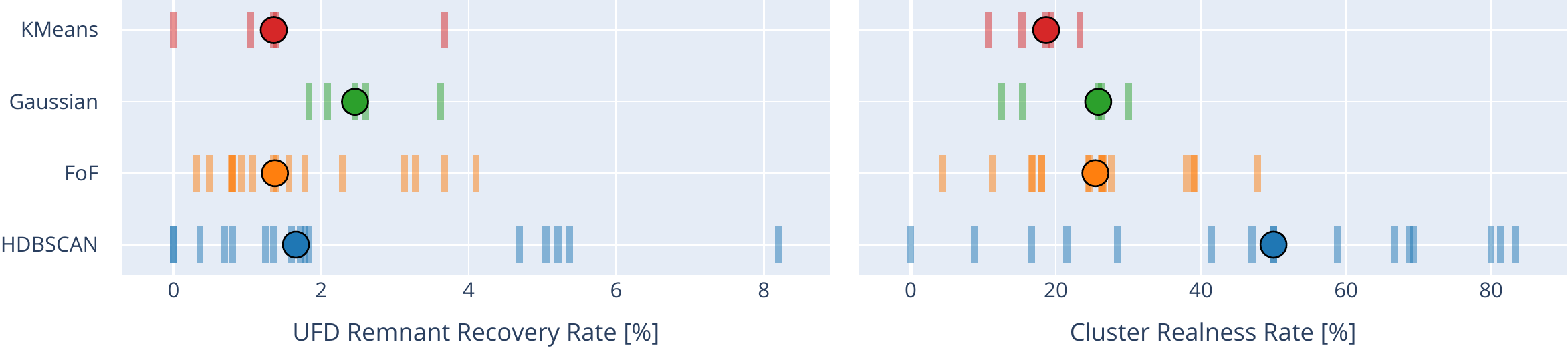}%
}

(c) All stars + entire halo data sets: HDBSCAN once again has the best balance of recovery and realness. \\[4ex]

\caption{Results for the all-stars data sets. Each line represents a single Milky Way-mass galaxy simulation and each circle is the median rate across all simulations. 
On the large data sets, HDBSCAN and FoF are much faster than K-means and Gaussian Mixture Models.
\label{fig:allalgos}}
\end{figure*}

\subsection{Comparing Clustering Algorithms}






Throughout this work, we test seven common clustering algorithms (described in Section \ref{subsec:algos}). For the UFD-only data sets, we test all seven algorithms on every data set. For the all-stars data sets, the larger radial cuts (50 kpc and entire halo) are extremely large, so we only test the more scalable algorithms: HDBSCAN, Friend-of-Friends, Gaussian Mixture Models, and K-Means.

The results for all UFD-only data sets are shown in Figure \ref{fig:UFDalgos}. Each line represents the results for a single Milky Way-mass galaxy simulation with the given radial cut. The median result for each algorithm is shown as a circle. The scatter in results across different simulations is significant because Milky Way-mass galaxies with a higher number of recent UFD accretions have higher rates. The results for the all-stars data sets are shown in Figure \ref{fig:allalgos}.

Even with UFD-only data sets, all algorithms have low UFD remnant recovery rates and cluster realness rates. The local radial cut, 5 kpc, has the worst results; the number of UFD remnants recovered from these simulations is frequently just one. \clearpage Overall, all algorithms only recover about 2\% of UFD remnants within 5 kpc of the Sun. HDBSCAN and FoF have the highest realness rates for the clusters they find, with around $20\%$ of their clusters corresponding to tagged star particles that accreted together.

This clearly implies, in no uncertain terms, that the vast majority of clusters found by these algorithms do not actually represent any truly accreted groups!

In the larger data sets, the clustering algorithms perform better, recovering $\sim3-6\%$ of UFD remnants and, for HDBSCAN and FoF, having a $\sim40-60\%$ realness rate. Even with these idealized data sets and specially chosen hyperparameters, though, the rates are still low. We thus discuss how to identify real clusters vs. false positives in Section \ref{sec:real}.

Generally for the UFD-only data sets, HDBSCAN is the most reliable algorithm choice. FoF also often has a relatively high realness rate. These two algorithms are also the fastest choices for large data sets.

For the all-stars data sets (Figure \ref{fig:allalgos}), realness rates are higher than the UFD-only data sets because clusters of stars from larger dwarf galaxy remnants are easier to identify than the small clusters of stars from UFD remnants. UFD remnant recovery rates are universally worse in the all-stars data sets, though, because the non-UFD stars act as significant noise during the search for UFD clusters. This is discussed in more detail in Section \ref{subsec:UFDvsall}.
Similar to the UFD-only data sets, HDBSCAN is once again generally a reliable choice to balance recovery rates and realness rates in the all-stars data sets. For the largest data sets, computational constraints also become important, and HDBSCAN and FoF scale well computationally.

Overall, HDBSCAN tends to be the most reliable clustering algorithm across different data sets. Currently, it is also a popular clustering algorithm used in astronomy research (see Section \ref{subsec:algos}). 
We thus focus on HDBSCAN for most of the rest of our text.

\subsection{Comparing UFD-Only Data Sets to All-Stars Data Sets} \label{subsec:UFDvsall}

\begin{figure}[tb]
\center
\includegraphics[width=0.45\textwidth]{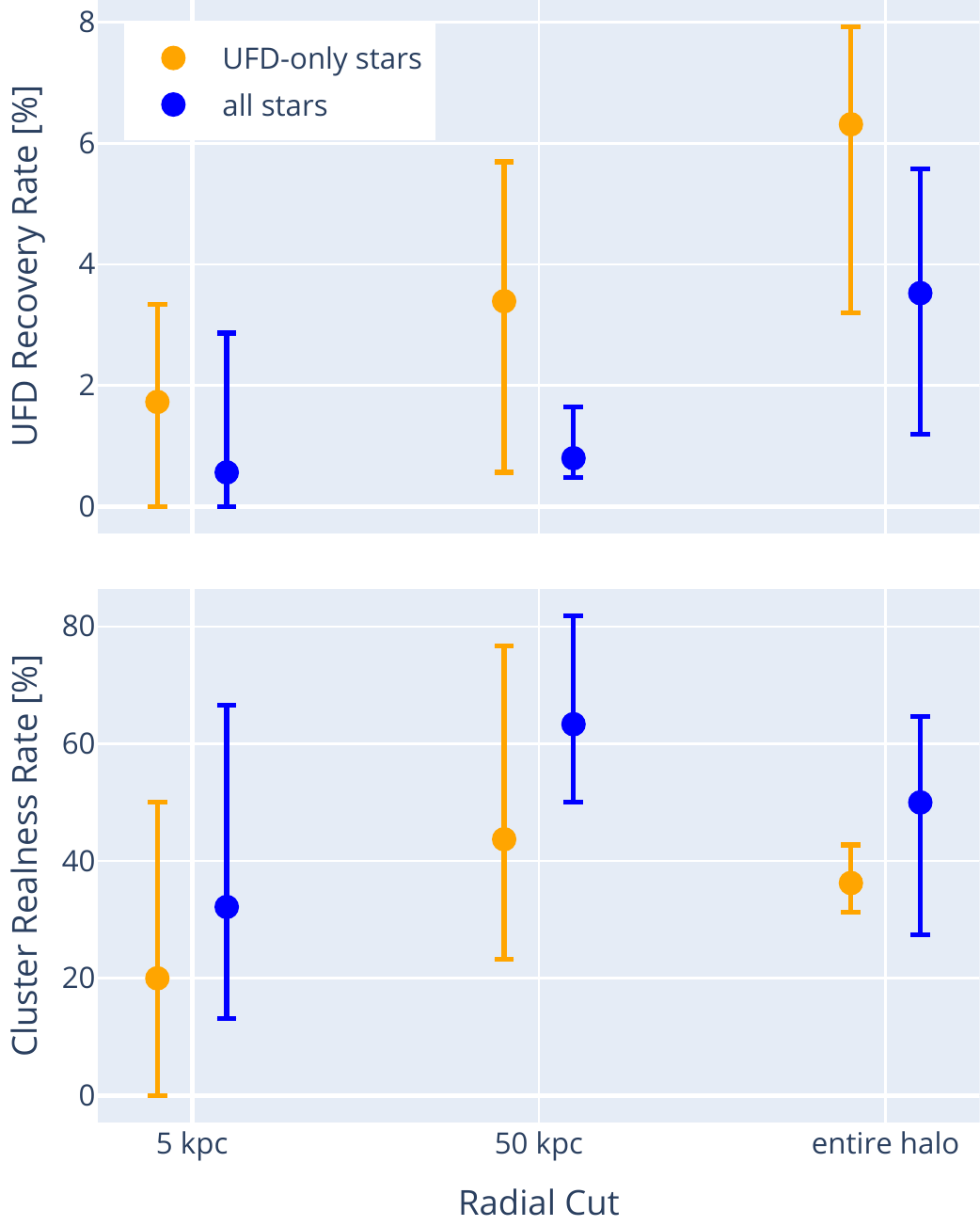}%

\caption{Comparing UFD remnant recovery rates and cluster realness rates with HDBSCAN for the data sets with only UFD stars and the data sets with all stars. As expected, the UFD-only data sets result in higher UFD recovery rates. Realness rates for the all-stars data sets include real clusters from larger dwarfs, which are principally easier to identify, so overall realness rates are not improved by using a UFD-only data set. Error bars show 16\%-84\% scatter across all simulations.
\label{fig:UFDvsall}}
\end{figure}

As discussed in Section \ref{subsec:datsets}, we have data sets with (1) only accreted star particles from UFDs and (2) all accreted star particles. The former data set is unrealistic because in real data we cannot know a priori which stars accreted from UFDs. The UFD-only data set can be imperfectly pursued observationally through the use of chemical tagging, however. Stars that formed in UFDs tend to have a lower metallicity distribution function, lower abundances in neutron-capture elements, and may preferentially have strong $r$-process enrichment \citep[e.g.,][]{Kirby13, Brauer19,Gudin21, Ji16}. Additionally, as we identify kinematic structures associated with larger-mass accretion events such as Gaia-Enceladus, removing those stars from observational data sets could also help towards creating a UFD-only data set. All these methods are imperfect, but as no more sophisticated and reliable methods exist to date to identify UFD stars e.g., in observed survey data, we must do the best we can with the methods available to us.

\begin{figure*}[tbh]
\center
\subfloat{
\includegraphics[width=0.95\textwidth]{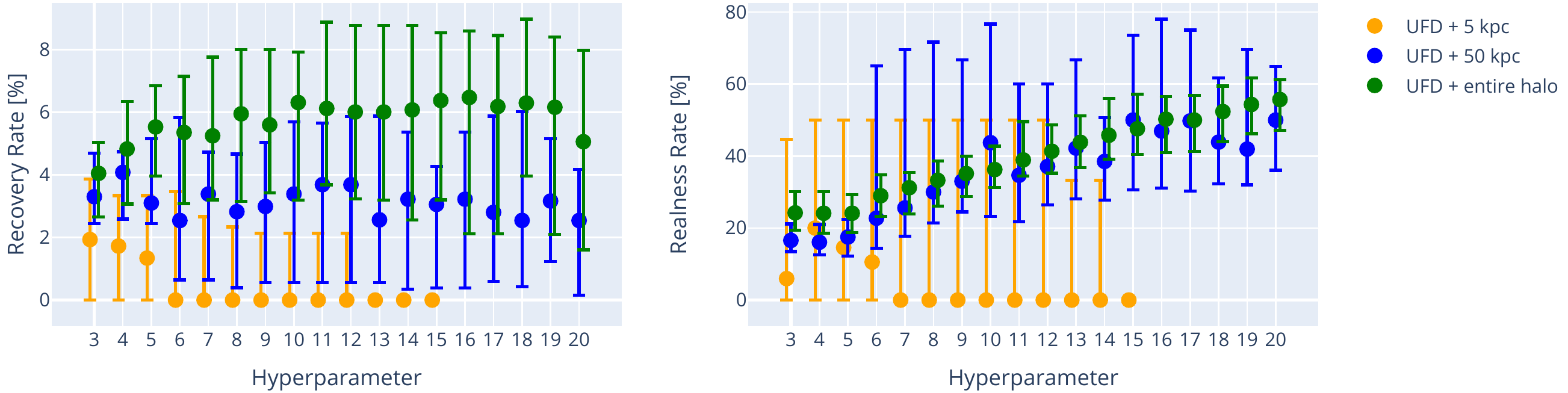}%
}

(a) Recovery rates and realness rates for different choices of min\_cluster\_size for HDBSCAN. \\[4ex]

\subfloat{
\includegraphics[width=0.95\textwidth]{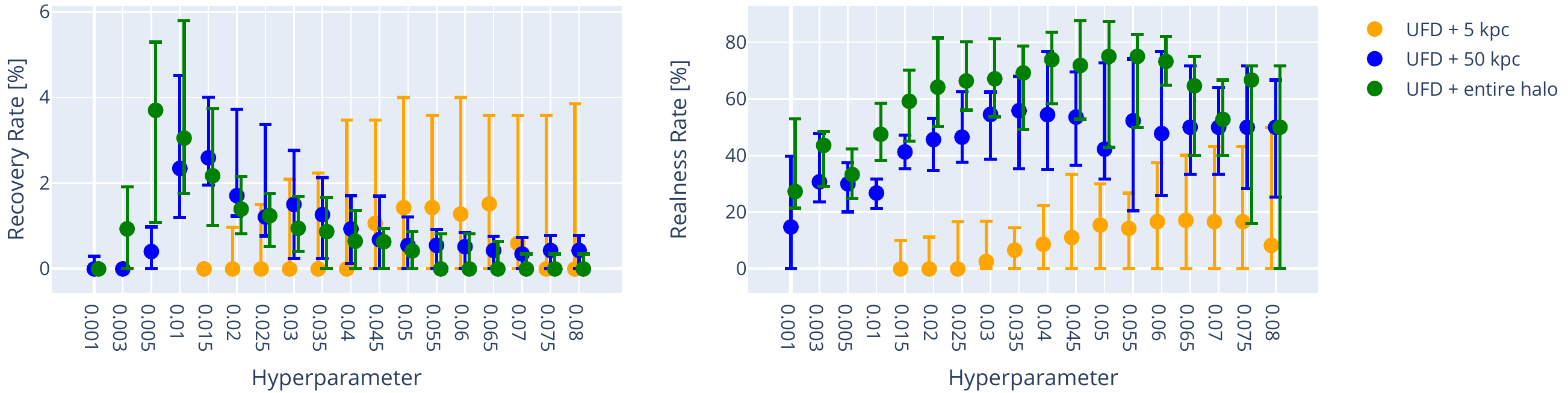}%
}

(b) Recovery rates for different choices of linking\_length for FoF. \\[2ex]

\caption{For these data sets, the FoF results differ more with varying hyperparameter choices than the HDBSCAN result. The hyperparameter choice is important for all algorithms, however. This causes additional difficulty when using these algorithms to identify UFD remnants.  Error bars show 16\%-84\% scatter across all simulations.
\label{fig:hyperparams}}
\end{figure*}

In Figure \ref{fig:UFDvsall}, we demonstrate the need to find ways to exclude stars from higher-mass accreted dwarfs if we hope to identify UFD remnants. 
At every radial cut, UFD remnant recovery rates are higher for UFD-only data sets. Realness rates are higher for all-stars data sets, but this is only because structures from higher-mass dwarfs are principally easier to identify than those from UFDs and because pure clusters are generally more common for higher-mass dwarfs since they contribute more stars. This underscores how difficult it is to identify UFD structures even among UFD-only samples. If we hope to identify UFD remnants, though, pursuing data sets with stars from UFDs will be, unsurprisingly, very beneficial.


\subsection{Comparing Hyperparameter Choices}

One downfall of most of these clustering algorithms is their dependence on hyperparameters. Each algorithm other than Affinity Propagation requires users to pre-select a value for a hyperparameter, and it is generally not obvious which values are best. In this work, we already know the true labels, and thus have the unique privilege of selecting our hyperparameters to optimize our clustering results (see Section \ref{subsec:hyperparam}). For observational data sets, however, this is not possible.

The results in all other subsections use optimal hyperparameter values. In this subsection, we vary the hyperparameter choices to illustrate how results differ. Figure \ref{fig:hyperparams} shows results for different hyperparameter choices of HDBSCAN and FoF. HDBSCAN requires an integer choice for min\_cluster\_size and thus has a smaller reasonable range of choices. Results can vary significantly with min\_cluster\_size choice, but generally results are roughly stable across several integer choices. As expected, the best choice of min\_cluster\_size tends to increase for data sets with larger radial cuts. For FoF, we tested many possible choices for linking\_length and results were more unstable than for HDBSCAN.

Thus, for these data sets, the results from HDBSCAN are more stable with variations in hyperparameter choice. The hyperparameter choice is important for all algorithms, however. This remains a difficulty of automating the search for UFD remnants with these clustering algorithms. Some groups are developing algorithms without a hyperparameter dependence \citep[e.g.,][]{RuizLara22} to alleviate these concerns.

Still, for HDBSCAN, the hyperparameter value greatly affects the number of clusters. For too large of min\_cluster\_size, the algorithm finds no remnants. For example, for the 5 kpc data sets, min\_cluster\_size $>5$ causes, on average, fewer than five total clusters returned by the algorithm, none of which are real UFD remnants. For the larger radial cuts, too small of min\_cluster\_size leads to too many clusters. For these data sets, min\_cluster\_size $<9$ causes 200 to 2000 clusters while the number of recovered remnants remains constant or decreases. When selecting this hyperparameter, a balance must be struck to avoid the identification of an unreasonably small or large number of clusters in a given sample.

\subsection{Why Clustering Algorithms Struggle} \label{subsec:struggle}

Due to their small size, the dynamic signatures of tidally-disrupted UFDs are, over 90\% of the time, weak and significantly out-numbered by other overlapping accreted structures. The limitations found in this paper are not unique to these clustering algorithms; we expect any clustering algorithm to struggle.

\begin{figure}[tb]
\center
\includegraphics[width=0.45\textwidth]{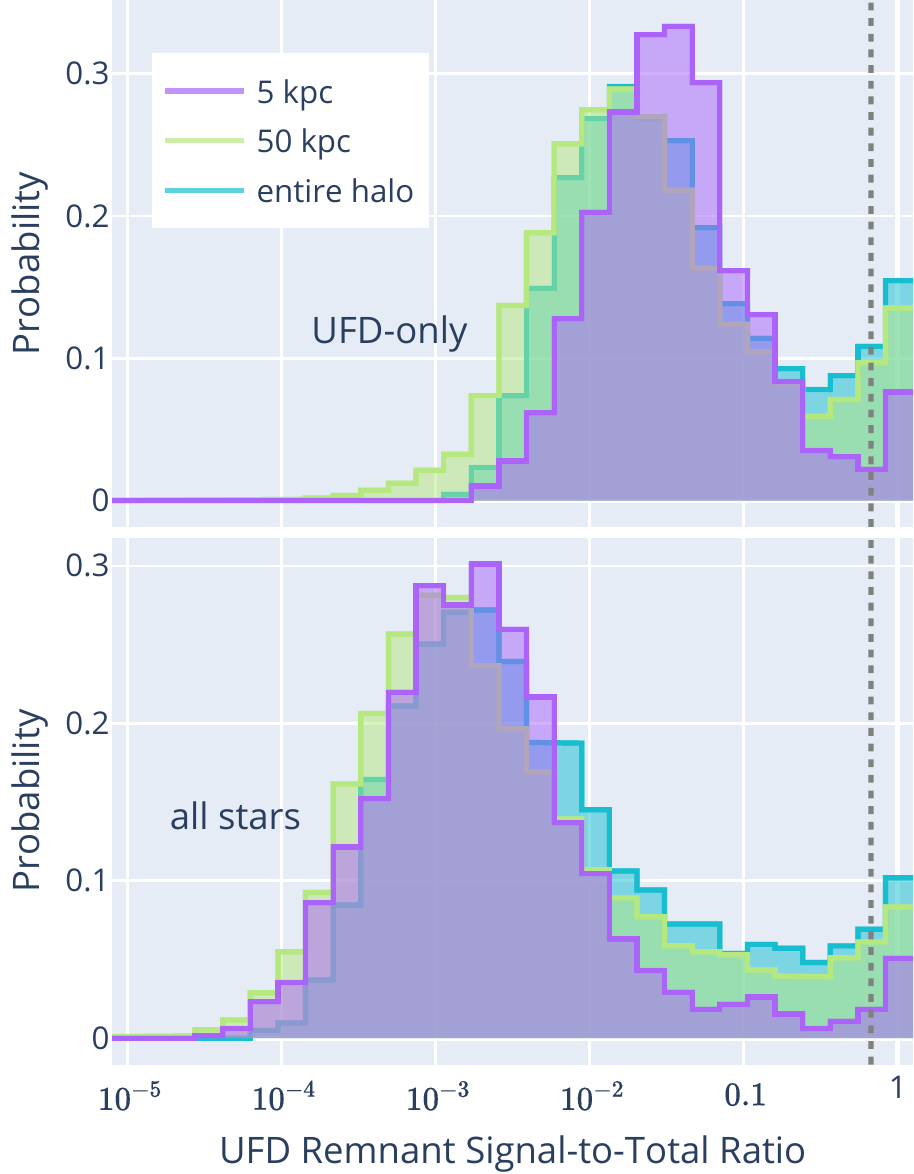}%
\caption{Normalized histograms of estimated signal-to-total ratios (similar to signal-to-noise ratios) for all UFD remnants in all data sets. For the vast majority of remnants, the ratio is tiny because the UFD remnant particles significantly overlap with all the other particles. The signal is very weak. For context, the dotted line shows 67\%, our purity threshold. Depending on the data set, $92-97$\% of UFD remnants have a signal-to-total ratio below this threshold. The all-stars data sets (bottom plot) have particularly low UFD signals -- the median ratio is one UFD remnant particle to 1000 non-remnant particles.
\label{fig:SNR}}
\end{figure}

To illustrate this, we estimate signal-to-total ratios (similar to signal-to-noise ratios) for all the tidally-disrupted UFD remnants in our data sets. Normalized histograms of the signal-to-total ratios from different data sets are shown in Figure \ref{fig:SNR}. To determine these ratios, for each remnant we draw a 4D sphere in phase space that is exactly large enough to enclose 50\% of the particles from that remnant. We then compare the number of remnant particles in that volume to the total number of particles in that volume. The maximum value is thus 1 for the case where the tidally-disrupted UFD is isolated from other particles. These ratios are similar to our purity metric, so we plot our purity threshold ($67\%$) as a dotted line on Figure \ref{fig:SNR} for reference. We also note that remnants are generally not spherical in 4D phase space, so this is merely an estimate.

For the vast majority of UFD remnants, the dynamic signature is completely washed out by the other particles in that volume. For UFD-only data sets, the typical remnant has a ratio of one UFD remnant particle to 30 other particles, 1:30. For the all-stars data sets, the typical remnant has a ratio of 1:1000. In the best case scenario, the UFD-only data set with the entire halo, only 8\% of remnants have a signal-to-total ratio higher than our purity threshold of 67\%.

The remnants with the highest signal-to-total ratios are the remnants that are successfully identified by the clustering algorithms. Most of the other remnants are simply too difficult to find in this dense 4D space, due to a combination of phase-mixing as the stellar dynamics relax over time and/or accreting with dynamics that are already similar to other star particles. We can thus optimize clustering searches to try to find the greatest number of UFD remnants, but most will never be found by these methods. The ones that are kinematically identifiable are those that (1) accreted with outlying dynamics, e.g., higher energy than usual, and (2) recently accreted so that the star particles have not had time to phase-mix.

We also note that an additional difficulty of analyzing only star particles in the inner volume, e.g. our 5 kpc data sets, is that you cannot sample full satellites within this small volume. This issue is described in more detail in \citet{Gomez10}.

\section{Properties of Real Clusters in Simulations} \label{sec:real}

Even in the best cases, the clustering algorithms find many clusters that do not correspond to real accreted remnant groups. Hence, we compare the properties of real clusters vs. ``false positive'' clusters to help inform which clusters are more likely to be real in observational data sets.

\begin{figure}[tb]
\center
\includegraphics[width=0.45\textwidth]{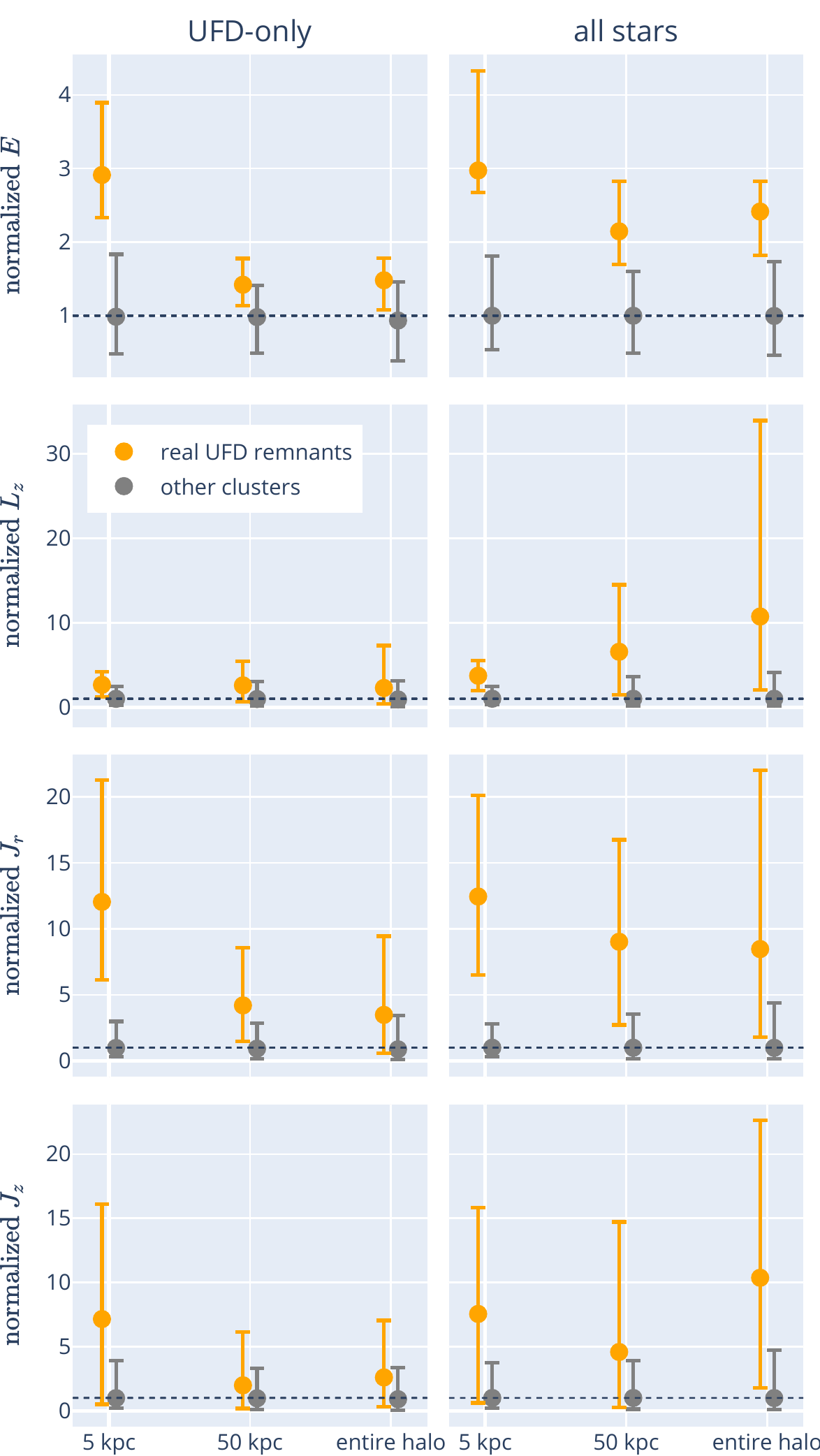}%
\caption{Median dynamics for clusters that correspond to real UFD remnants (i.e., pure and complete clusters) compared to other clusters. Clusters with higher actions are more likely to be real.  Error bars show 16\%-84\% scatter across all clusters.
\label{fig:recoveractions}}
\end{figure}

Figure \ref{fig:recoveractions} shows the $E$, $L_z$, $J_r$, and $J_z$ of real recovered clusters (i.e., pure and complete clusters -- clusters that correspond to an accreted UFD remnant) compared to the dynamics of clusters that do not correspond to UFD remnants. These results use HDBSCAN, but the plots are holistically similar for other algorithms. All dynamics are normalized relative to the median of all clusters in the sample. For each cluster, its energy (or $L_z$, $J_r$, $J_z$) is determined from the median of all star particles in that cluster.

Compared to all clusters, clusters that correspond to real UFD remnants have higher energy and axisymmetric actions. High energy and $J_r$ are most important for distinguishing between real UFD clusters and all other clusters, especially in local (5 kpc) data sets. Of the action variables, $L_z$ is the least important dynamic when determining which clusters are more likely to be real. This aligns with results from the ANOVA tests in Figure \ref{fig:omegasq}.

Based on these results, clusters with high energy and high $J_r$ are significantly more trustworthy. For example, clusters with median energy higher than twice the median of all clusters in a local sample are pure \textit{and} complete over 90\% of the time. This is true for both UFD-only data sets and all-stars data sets.

\begin{figure}[tb]
\center
\includegraphics[width=0.47\textwidth]{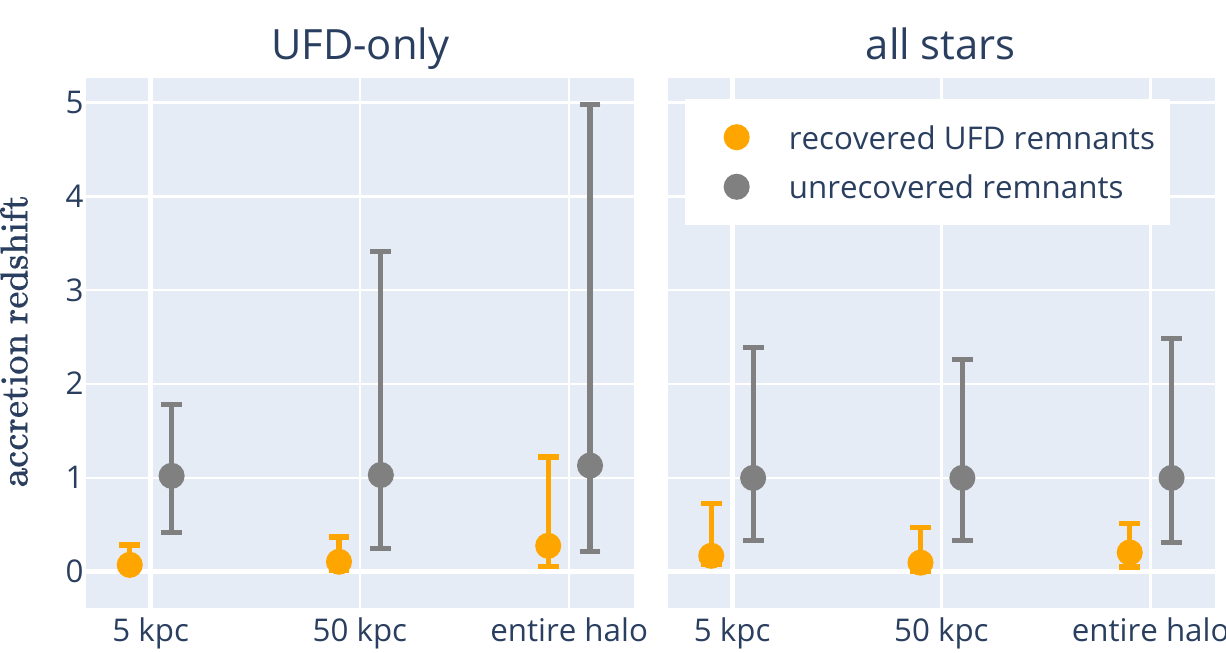}%
\caption{Median $z_{accretion}$ (redshift at which a given dwarf galaxy was accreted) for recovered UFD remnants compared to all unrecovered remnants. As expected, the UFD remnants that are recovered by HDBSCAN (and other algorithms) were more recently accreted.  Error bars show 16\%-84\% scatter across all remnants.
\label{fig:recoverz}}
\end{figure}

The UFD remnants recovered in these real clusters are UFDs that, generally, accreted relatively recently. Figure \ref{fig:recoverz} shows the median accretion redshift $z_{accretion}$ for UFDs recovered by HDBSCAN compared to all unrecovered remnants. UFDs that were accreted at redshift $z=1$ and higher are virtually never recovered by any of these clustering algorithms. The dynamic signature of these small dwarfs is completely lost as the stars phase-mix in the dense region of action space, and the remnants are no longer identifiable. This is not surprising because energy and orbital actions are only truly conserved in static potentials, and realistic, time-varying galactic potentials cause the stellar dynamics to relax over time.

As discussed in Section \ref{subsec:struggle}, for a UFD remnant to be reliably identified through kinematic clustering, it needs to both have had outlying dynamics at the time of accretion and also have a recent accretion time, $z_{accretion} \lesssim 0.5$, so that its stars have not had time to significantly phase-mix. Not all recently accreted UFD remnants are identifiable through kinematics (recently accreted UFDs can still end up in the dense regions of phase space; see Figure \ref{fig:infall}), but of the identifiable UFD remnants, virtually all are recently accreted.

\section{Recommendations for Using Cluster Algorithms}
\label{sec:takeaways}

Our study has clearly shown that using clustering algorithms with stellar dynamics to search for accreted UFD remnants is a challenging task, that, unfortunately, does not deliver reliable results a majority of the time.

Dynamically-linked clusters identified by any clustering algorithms should thus not be blindly trusted but amply questioned and investigated, and results presented in a careful manner to avoid the presentation of numerically artificially created results. Case in point is our idealized situations in which we limit our data sets to only accreted UFD star particles and optimize our hyperparameter choices. The resulting UFD recovery rates are around $\sim6\%$ at best, and the majority of clusters found by all algorithms are not real. Only stars from fairly recently accreted UFDs ($z_{accretion} \lesssim 0.5$) can retain sufficiently strong dynamic signatures to be identified by these algorithms.

While these findings are unfortunate and must be taken into account in future searches, not all is lost. 
Clustering with stellar dynamics remains one of the few methods presently available to identify accreted structure in observed Milky Way survey data, and while not all UFDs can be found this way, identifying real remnants is possible. 

To ensure that results are as reliable and trustworthy as possible, we recommend that researchers:




\begin{itemize}
    \item Among these out-of-the-box clustering algorithms, choose HDBSCAN. Across our different data sets, HDBSCAN consistently balances the highest UFD remnant recovery rates and cluster realness rates. It is also more computationally scalable than all algorithms other than Friend-of-Friends.
    \item UFD dynamic signatures are frequently weak, so incorporate chemical tagging when identifying groups of accreted stars. This can be done, for example, by focusing on low-metallicity stars and/or $r$-process enhanced stars. Successfully limiting a data set to UFD stars increases your remnant recovery rate by around $3\times$ on average. Chemical abundances can also be used to help validate dynamic clusters.
    \item Assume most clusters identified by clustering algorithms do not correspond to real UFD remnants. Focus on clusters with higher than average energy and $J_r$.
    \item Recognize that only recently accreted UFDs in lower-density areas of phase space are consistently found by these clustering algorithms, so you generally only recover $1-6\%$ of the UFD remnants in a given sample. Samples limited to the region around the Sun have lower recovery rates than samples with larger radial cuts.
    \item Vary your hyperparameter choices and consider the stability of the clustering results across several hyperparameter values. For HDBSCAN, the best hyperparameter values are the ones which produce fewer than several hundred clusters (in our samples, requires $\text{min\_cluster\_size} \gtrsim 9$ for our large radial cuts) and produce more than just a few clusters (in our samples, requires $\text{min\_cluster\_size} \lesssim 6$ for our 5 kpc radial cut). This will depend on your sample, so test different hyperparameter choices to avoid hyperparameters that result in an unreasonably large or small number of clusters.
\end{itemize}

\begin{acknowledgments}
KB acknowledges support from the United States Department of Energy grant DE-SC0019323. AF acknowledge support from NSF grant AST-1716251. FAG acknowledges support from ANID FONDECYT Regular 1211370 and by the ANID BASAL project FB210003. FAG also acknowledges funding from the Max Planck Society through a “Partner Group” grant.

This work made extensive use of the python libraries \texttt{numpy} \citep{numpy}, \texttt{scipy} \citep{2020SciPy-NMeth}, \texttt{sqlite3} \citep{sqlite3}, and \texttt{plotly} \citep{plotly}.
\end{acknowledgments}

\vspace{5mm}


\appendix

\begin{table*}[!htb]
\centering
\caption{Results for the one-way ANOVA tests. Each continuous variable in the table is tested for its level of association to the true cluster labels. The $\omega^2$ values estimate the strength of the association; a high $\omega^2$ value (e.g., near 1) implies that this variable is likely to be important in clustering. Uncertainty values represent 16th -- 84th percentile scatter across the 32 simulations. 
} \label{tab:ANOVA}
\begin{tabular}{|c|c|c|c|c||c|c|c|c|c|}
\tablewidth{0pt}
\hline
\hline
Variable & Radial Cut & $F$ & $\omega^2$ & $p-$value & Variable & Radial Cut & $F$ & $\omega^2$  & $p-$value  \\
\hline
 & 5 kpc & $2^{+1}_{-1}$ & $0.14^{+0.04}_{-0.05}$ & $2\text{e-}10^{+6\text{e-}8}_{-2\text{e-}10}$ &  & 5 kpc & $2^{+1}_{-1}$ & $0.12^{+0.05}_{-0.04}$ & $6\text{e-}11^{+3\text{e-}4}_{-6\text{e-}11}$ \\
$\rho$ & 50 kpc & $186^{+93}_{-45}$ & $0.38^{+0.10}_{-0.06}$ & $<1\text{e-}300$ & $L_r$ & 50 kpc & $28^{+23}_{-7}$ & $0.09^{+0.04}_{-0.03}$ & $<1\text{e-}300$ \\
 & entire halo & $460^{+407}_{-97}$ & $0.56^{+0.13}_{-0.07}$ & $<1\text{e-}300$ & & entire halo & $71^{+75}_{-21}$ & $0.15^{+0.13}_{-0.04}$ & $<1\text{e-}300$ \\
\hline
 & 5 kpc & $1^{+1}_{-1}$ & $0.02^{+0.06}_{-0.02}$ & $3\text{e-}2^{+4\text{e-}1}_{-3\text{e-}2}$  & & 5 kpc & $4^{+1}_{-1}$ & $0.21^{+0.09}_{-0.07}$ & $2\text{e-}25^{+3\text{e-}13}_{-2\text{e-}25}$ \\
$\phi$ & 50 kpc & $13^{+4}_{-3}$ & $0.04^{+0.01}_{-0.01}$ & $<1\text{e-}300$ & $L_\phi$ & 50 kpc & $31^{+20}_{-7}$ & $0.10^{+0.04}_{-0.02}$ & $<1\text{e-}300$ \\
 & entire halo & $33^{+13}_{-5}$ & $0.07^{+0.03}_{-0.01}$ & $<1\text{e-}300$ & & entire halo & $98^{+158}_{-45}$ & $0.19^{+0.20}_{-0.08}$ & $<1\text{e-}300$ \\
\hline
 & 5 kpc & $2^{+1}_{-1}$ & $0.05^{+0.03}_{-0.03}$ & $7\text{e-}4^{+2\text{e-}1}_{-7\text{e-}4}$ &  & 5 kpc & $3^{+2}_{-1}$ & $0.20^{+0.08}_{-0.08}$ & $2\text{e-}25^{+2\text{e-}8}_{-2\text{e-}25}$ \\
$z$ & 50 kpc & $27^{+9}_{-6}$ & $0.08^{+0.03}_{-0.01}$ & $<1\text{e-}300$ & $L_z$ & 50 kpc  & $110^{+101}_{-41}$ & $0.28^{+0.13}_{-0.09}$ & $<1\text{e-}300$ \\
 & entire halo & $161^{+147}_{-61}$ & $0.31^{+0.14}_{-0.10}$  & $<1\text{e-}300$ & & entire halo & $243^{+170}_{-101}$ & $0.40^{+0.11}_{-0.14}$ & $<1\text{e-}300$ \\
\hline
 & 5 kpc & $2^{+2}_{-1}$ & $0.15^{+0.04}_{-0.05}$ & $1\text{e-}10^{+3\text{e-}8}_{-1\text{e-}10}$ & & 5 kpc & $10^{+9}_{-3}$ & $0.51^{+0.06}_{-0.07}$ & $2\text{e-}93^{+1\text{e-}42}_{-2\text{e-}93}$ \\
$r$ & 50 kpc & $244^{+137}_{-50}$ & $0.45^{+0.13}_{-0.05}$ & $<1\text{e-}300$ & $L_{total}$ & 50 kpc & $392^{+111}_{-69}$ & $0.58^{+0.04}_{-0.07}$ & $<1\text{e-}300$ \\
 & entire halo & $525^{+466}_{-63}$ & $0.59^{+0.12}_{-0.07}$ & $<1\text{e-}300$ & & entire halo & $600^{+382}_{-198}$ & $0.60^{+0.11}_{-0.10}$ & $<1\text{e-}300$ \\
\hline
 & 5 kpc & $2^{+1}_{-1}$ & $0.14^{+0.05}_{-0.05}$ & $2\text{e-}11^{+1\text{e-}6}_{-2\text{e-}11}$  & & 5 kpc & $22^{+30}_{-12}$ & $0.74^{+0.08}_{-0.25}$ & $3\text{e-}166^{+6\text{e-}72}_{-3\text{e-}166}$ \\
$v_r$ & 50 kpc & $15^{+9}_{-4}$ & $0.05^{+0.03}_{-0.01}$ & $<1\text{e-}300$ & $J_r$ & 50 kpc & $373^{+360}_{-122}$ & $0.57^{+0.13}_{-0.09}$ & $<1\text{e-}300$ \\
 & entire halo & $28^{+10}_{-9}$ & $0.07^{+0.02}_{-0.02}$ & $<1\text{e-}300$ & & entire halo & $760^{+565}_{-298}$ & $0.66^{+0.10}_{-0.14}$ & $<1\text{e-}300$ \\
\hline
 & 5 kpc & $2^{+1}_{-1}$ & $0.11^{+0.06}_{-0.04}$ & $2\text{e-}13^{+3\text{e-}4}_{-2\text{e-}13}$  & & 5 kpc & $7^{+8}_{-3}$ & $0.40^{+0.13}_{-0.16}$ & $4\text{e-}55^{+1\text{e-}26}_{-4\text{e-}55}$ \\
$v_\phi$ & 50 kpc & $1^{+1}_{-1}$ & $0.00^{+0.00}_{-0.00}$ & $3\text{e-}4^{+10\text{e-}1}_{-3\text{e-}4}$  & $J_z$ & 50 kpc & $229^{+61}_{-68}$ & $0.44^{+0.07}_{-0.08}$  & $<1\text{e-}300$ \\
 & entire halo & $2^{+1}_{-1}$ & $0.00^{+0.00}_{-0.00}$ & $7\text{e-}7^{+9\text{e-}1}_{-7\text{e-}7}$ & & entire halo & $362^{+245}_{-98}$ & $0.49^{+0.16}_{-0.10}$ & $<1\text{e-}300$ \\
\hline
 & 5 kpc & $3^{+2}_{-1}$ & $0.19^{+0.09}_{-0.07}$  & $4\text{e-}21^{+1\text{e-}8}_{-4\text{e-}21}$ & & 5 kpc & $22^{+11}_{-9}$ & $0.68^{+0.07}_{-0.13}$ & $6\text{e-}156^{+2\text{e-}77}_{-6\text{e-}156}$ \\
$v_z$ & 50 kpc & $35^{+15}_{-7}$ & $0.10^{+0.04}_{-0.02}$ & $<1\text{e-}300$ & $E$ & 50 kpc & $934^{+755}_{-319}$ & $0.77^{+0.07}_{-0.08}$ & $<1\text{e-}300$ \\
 & entire halo & $43^{+14}_{-8}$ & $0.09^{+0.03}_{-0.01}$ & $<1\text{e-}300$ & & entire halo & $1489^{+1031}_{-567}$ & $0.80^{+0.07}_{-0.10}$ & $<1\text{e-}300$\\
\hline
 & 5 kpc & $12^{+7}_{-4}$ & $0.56^{+0.06}_{-0.12}$ & $2\text{e-}98^{+5\text{e-}51}_{-2\text{e-}98}$  & & 5 kpc & $(6^{+11}_{-5})\times 10^{27}$ & 1.00$^{+0.00}_{-0.00}$   & $<1\text{e-}300$  \\
$v_{total}$ & 50 kpc & $117^{+39}_{-37}$ & $0.28^{+0.07}_{-0.07}$ & $<1\text{e-}300$ & $z_{infall}$ & 50 kpc & $(5^{+5}_{-4})\times 10^{27}$ & 1.00$^{+0.00}_{-0.00}$ & $<1\text{e-}300$\\
 & entire halo & $67^{+38}_{-22}$ & $0.14^{+0.06}_{-0.03}$ & $<1\text{e-}300$ & & entire halo & $(4^{+3}_{-3})\times 10^{27}$ & 1.00$^{+0.00}_{-0.00}$ & $<1\text{e-}300$ \\ 
\hline
\end{tabular} 
\end{table*}


\begin{table}[]
\caption{Trial hyperparameter values for all algorithms.}
\begin{tabular}{|c|c|c|}
\hline
\hline
Algorithm  & Hyperparameter & Hyperparameter Search Space \\         \hline
HDBSCAN  & min\_cluster\_size & 3, 4, 5, 6, 7, 8, 9, 10, 11, 12, \\
& & 13, 14, 15, 16, 17, 18, 19, 20 \\ 
\hline
K-means & n\_clusters & 60, 70, 80, 90, 100, 110, 120,  \\ 
\cline{1-2}
Gaussian mixture models  & n\_clusters & 130, 140, 150, 160, 170, 180, 190, \\ 
\cline{1-2}
Agglomerative clustering & n\_clusters & 200, 210, 220, 230, 240, 250\\ 
\hline
& & 0.001, 0.003, 0.005, 0.01, 0.015, 0.02, 0.025,  \\
Friend-of-friends & linking\_length &  0.03, 0.035, 0.04, 0.045, 0.05, 0.055, 0.06, 0.065,  \\ 
& &  0.07, 0.075, 0.08, 0.1, 0.125, 0.15, 0.175, 0.2 \\ 
\hline
Mean shift & bandwidth & 0.06, 0.07, 0.08, 0.09, 0.1, 0.11, 0.12, 0.13,\\
& & 0.14, 0.15, 0.16, 0.17, 0.18, 0.19, 0.2 \\ 
\hline
Affinity propagation & - & - \\ 
\hline
\end{tabular}
\end{table}

\clearpage

\begin{table}[]
\center
\caption{Optimal hyperparameter values for all datasets}
\begin{tabular}{|c|c|cccccc|}
\hline
\multirow{2}{*}{} & \multirow{2}{*}{} & \multicolumn{6}{c|}{Chosen Hyperparameters} \\ \cline{3-8} 
&  & \multicolumn{3}{c|}{only UFD particles}  & \multicolumn{3}{c|}{UFD and non-UFD particles} \\ 
\hline
Algorithm & Hyperparameter & \multicolumn{1}{c|}{5 kpc} & \multicolumn{1}{c|}{50 kpc} & \multicolumn{1}{c|}{no radial cut} & \multicolumn{1}{c|}{5 kpc} & \multicolumn{1}{c|}{50 kpc} & no radial cut \\ \hline

HDBSCAN & min\_cluster\_size & \multicolumn{1}{c|}{4} & \multicolumn{1}{c|}{10}      & \multicolumn{1}{c|}{15}             & \multicolumn{1}{c|}{3}     & \multicolumn{1}{c|}{10}     & 19            \\ \hline
K-means                    & n\_clusters                     & \multicolumn{1}{c|}{60}   & \multicolumn{1}{c|}{230}    & \multicolumn{1}{c|}{180}           & \multicolumn{1}{c|}{200}    & \multicolumn{1}{c|}{230}    & 150            \\ \hline
Gaussian mixture models    & n\_clusters                     & \multicolumn{1}{c|}{90}    & \multicolumn{1}{c|}{160}    & \multicolumn{1}{c|}{210}           & \multicolumn{1}{c|}{70}    & \multicolumn{1}{c|}{220}     & 240           \\ \hline
Agglomerative clustering   & n\_clusters                     & \multicolumn{1}{c|}{80}    & \multicolumn{1}{c|}{250}    & \multicolumn{1}{c|}{250}           & \multicolumn{1}{c|}{170}    & \multicolumn{1}{c|}{too slow}    & too slow            \\ \hline
Friend-of-friends         & linking\_length                 & \multicolumn{1}{c|}{0.065}  & \multicolumn{1}{c|}{0.015}  & \multicolumn{1}{c|}{0.01}          & \multicolumn{1}{c|}{0.065} & \multicolumn{1}{c|}{0.015}    & 0.01          \\ \hline
Mean shift                 & bandwidth                       & \multicolumn{1}{c|}{0.11}   & \multicolumn{1}{c|}{0.06}   & \multicolumn{1}{c|}{0.06}         & \multicolumn{1}{c|}{0.16}  & \multicolumn{1}{c|}{too slow}   & too slow        \\ \hline
Affinity propagation       & -                               & \multicolumn{1}{c|}{-}     & \multicolumn{1}{c|}{-}      & \multicolumn{1}{c|}{-}             & \multicolumn{1}{c|}{-}     & \multicolumn{1}{c|}{-}      & -             \\ \hline
\end{tabular}
\end{table}

\bibliography{kaleybib}{}
\bibliographystyle{aasjournal}

\end{document}